\documentclass[aps,prl,superscriptaddress,amsmath,amssymb,floatfix,reprint]{revtex4-2}
\pdfoutput=1

\usepackage{amssymb}
\usepackage{graphicx}
\usepackage{bm}
\usepackage[pdftex,breaklinks=true,bookmarksopen=true,bookmarksopenlevel=3,bookmarksnumbered=true,colorlinks=true,urlcolor= magenta,citecolor=blue,linkcolor=blue]{hyperref}
\usepackage{float}
\usepackage[utf8]{inputenc}
\usepackage[T1]{fontenc}
\usepackage{verbatim}
\usepackage{siunitx}
\usepackage[dvipsnames]{xcolor}
\usepackage{soul}
\usepackage{ulem}

\usepackage{ragged2e}

\usepackage{newfloat}
\usepackage{caption}
\DeclareFloatingEnvironment{smfigure}
\begin{document}

\title{Strong-field focusing of high-energy particles in beam-multifoil collisions}

\author{A. Matheron}

\email[Corresponding author: ]{a.matheron@hi-jena.gsi.de}

\altaffiliation[\\Present addresses: ]{Helmholtz-Institut Jena, Fr\"obelstieg 3, 07743 Jena, Germany; GSI Helmholtzzentrum f\"ur Schwerionenforschung, Planckstra\ss e 1, 64291 Darmstadt, Germany; Faculty of Physics and Astronomy, Friedrich-Schiller-Universit\"at Jena, 07743 Jena, Germany}

\affiliation{Laboratoire d’Optique Appliquée (LOA), CNRS, École polytechnique, ENSTA, Institut Polytechnique de Paris, Palaiseau, France}

\author{D. Storey}
\email[Corresponding author: ]{dstorey@slac.stanford.edu}
\affiliation{SLAC National Accelerator Laboratory, Menlo Park, CA 94025, USA}
\author{M. F. Gilljohann}
\affiliation{Laboratoire d’Optique Appliquée (LOA), CNRS, École polytechnique, ENSTA, Institut Polytechnique de Paris, Palaiseau, France}
\author{S. Rego}
\affiliation{Laboratoire d’Optique Appliquée (LOA), CNRS, École polytechnique, ENSTA, Institut Polytechnique de Paris, Palaiseau, France}

\author{E. Adli}
\affiliation{Department of Physics, University of Oslo, 0316 Oslo, Norway}

\author{I. A. Andriyash}
\affiliation{Laboratoire d’Optique Appliquée (LOA), CNRS, École polytechnique, ENSTA, Institut Polytechnique de Paris, Palaiseau, France}

\author{G.~J.~Cao}
\affiliation{Department of Physics, University of Oslo, 0316 Oslo, Norway}

\author{X. Davoine}
\affiliation{CEA, DAM, DIF, F-91297 Arpajon, France}
\affiliation{Universit\'{e} Paris-Saclay, CEA, LMCE, F-91680 Bruy\`{e}res-le-Ch\^{a}tel, France}

\author{C. Emma}
\affiliation{SLAC National Accelerator Laboratory, Menlo Park, CA 94025, USA}

\author{F. Fiuza}
\affiliation{GPA/Instituto de Plasmas e Fusão Nuclear, Instituto Superior Técnico, Universidade de Lisboa, Lisbon, 1049-001, Portugal}

\author{S. Gessner}
\affiliation{SLAC National Accelerator Laboratory, Menlo Park, CA 94025, USA}

\author{L.~Gremillet}
\affiliation{CEA, DAM, DIF, F-91297 Arpajon, France}
\affiliation{Universit\'{e} Paris-Saclay, CEA, LMCE, F-91680 Bruy\`{e}res-le-Ch\^{a}tel, France}

\author{C. Hansel}
\affiliation{University of Colorado Boulder, Department of Physics, Center for Integrated Plasma Studies, Boulder, Colorado 80309, USA}

\author{C. Joshi}
\affiliation{University of California Los Angeles, Los Angeles, CA 90095, USA}

\author{C.~H.~Keitel}
\affiliation{Max-Planck-Institut f\"ur Kernphysik, Saupfercheckweg 1, D-69117 Heidelberg, Germany}

\author{A. Knetsch}
\affiliation{SLAC National Accelerator Laboratory, Menlo Park, CA 94025, USA}

\author{V. Lee}
\affiliation{University of Colorado Boulder, Department of Physics, Center for Integrated Plasma Studies, Boulder, Colorado 80309, USA}

\author{M. D. Litos}
\affiliation{University of Colorado Boulder, Department of Physics, Center for Integrated Plasma Studies, Boulder, Colorado 80309, USA}

\author{N. Majernik}
\affiliation{SLAC National Accelerator Laboratory, Menlo Park, CA 94025, USA}

\author{Y. Mankovska}
\affiliation{Laboratoire d’Optique Appliquée (LOA), CNRS, École polytechnique, ENSTA, Institut Polytechnique de Paris, Palaiseau, France}

\author{B. O'Shea}
\affiliation{SLAC National Accelerator Laboratory, Menlo Park, CA 94025, USA}

\author{I. Rajkovic}
\affiliation{SLAC National Accelerator Laboratory, Menlo Park, CA 94025, USA}

\author{P.~San~Miguel~Claveria}
\altaffiliation[Present address: ]{Instituto de Fusión Nuclear ``Guillermo Velarde'', Universidad Politécnica de Madrid, José Gutiérrez Abascal 2, 28006 Madrid, Spain}
\affiliation{Laboratoire d’Optique Appliquée (LOA), CNRS, École polytechnique, ENSTA, Institut Polytechnique de Paris, Palaiseau, France}

\author{V. Zakharova}
\affiliation{Laboratoire d’Optique Appliquée (LOA), CNRS, École polytechnique, ENSTA, Institut Polytechnique de Paris, Palaiseau, France}

\author{C. Zhang}
\affiliation{University of California Los Angeles, Los Angeles, CA 90095, USA}

\author{M. J. Hogan}
\affiliation{SLAC National Accelerator Laboratory, Menlo Park, CA 94025, USA}

\author{M.~Tamburini}
\email[Corresponding author: ]{\\matteo.tamburini@mpi-hd.mpg.de}
\affiliation{Max-Planck-Institut f\"ur Kernphysik, Saupfercheckweg 1, D-69117 Heidelberg, Germany}

\author{S.~Corde}
\email[Corresponding author: ]{\\sebastien.corde@polytechnique.edu}
\affiliation{Laboratoire d’Optique Appliquée (LOA), CNRS, École polytechnique, ENSTA, Institut Polytechnique de Paris, Palaiseau, France}

\begin{abstract}

Extreme beams of charged particles and photons, reaching ultrahigh densities or producing intense gamma-ray bursts, are central to accelerator physics, laboratory astrophysics, and strong-field quantum electrodynamics research. Yet their generation is hindered by conventional focusing methods at multi-GeV energies that rely on massive magnetic assemblies, limiting compactness and attainable density. Here we report the first experimental observation of a fundamentally new focusing mechanism, in which a high-energy charged-particle beam is focused by its own magnetic field reflected from a stack of thin metallic foils via near-field coherent-transition-radiation. The experiment, performed at SLAC’s FACET-II facility, reveals strong, cumulative focusing across a broad range of beam configurations, enabled by the delivered 10 GeV, 1 nC, 10 Hz electron beam. The measurements closely agree with predictions from an analytical model and particle-in-cell simulations. These results demonstrate that multifoil focusing is a remarkably straightforward, self-aligned approach to the generation of ultrahigh density beams, opening a path to explore unprecedented regimes of beam-matter interaction and high-energy radiation.

\end{abstract}

\maketitle

\begin{figure*}[t!] 
  \centering
  \includegraphics[width=18cm]{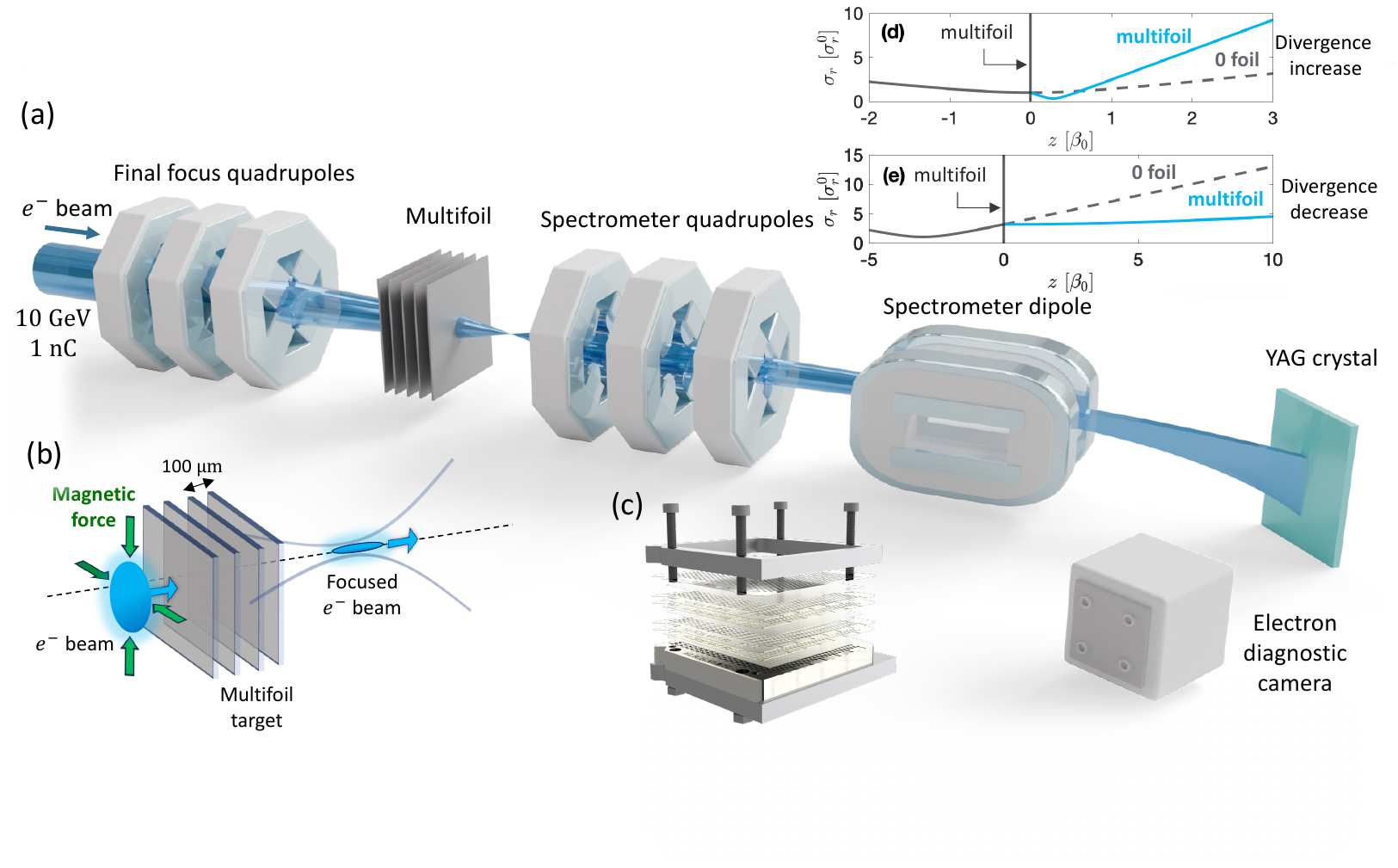}
  \vspace{-2cm}
  \caption{\justifying \textbf{Scheme for multifoil-induced beam focusing.} (a) Experimental setup at SLAC's FACET-II accelerator facility.  (b) Beam colliding with a multifoil and undergoing strong-field focusing. (c) 3D view of the multifoil target. (d-e) Evolution of the beam’s transverse size while traversing the multifoil target, as predicted by analytical theory, for the cases where the vacuum beam waist is located (d) at the target position and (e) upstream of the target. The beam size $\sigma_r$ is expressed in units of the beam size at the vacuum waist $\sigma_{r,0}$, while the propagation coordinate $z$ is normalized to the vacuum beta function at the waist $\beta_0$. The dashed line indicates the beam envelope for a vacuum drift.}
  \label{FIGURE_1}
\end{figure*}

The generation of charged particle beams of ultrahigh energies (> GeV) and ultrahigh densities ($>10^{21}\,\rm cm^{-3}$) is a prerequisite for frontier research in extreme radiation-matter interactions at the heart of relativistic plasma physics, laboratory astrophysics, and strong-field quantum electrodynamics (QED)~\cite{Bilderback2005,MourouExtremeBeams2006,Marklund2006,Ruffini2010,DiPiazza2012,Corde2013,Ullrich2012,UzdenskiArXiv2019,Xu2021}. However, producing such extreme beams remains technologically challenging. Conventional focusing relies on massive magnetic elements that become increasingly large and complex at high energies and, much like active plasma lenses, are ultimately limited in focusing power, as denser beams demand increasing strong focusing fields~\cite{Telnov1990}. Therefore, the generation of ultradense particle beams--and consequently high-flux gamma-ray bursts--remains difficult, limiting progress in forefront applications in multiple fields.

Here we experimentally demonstrate a strikingly simple new focusing mechanism: focusing a high-energy beam with nothing more than a few thin aluminum foils. By reflecting the beam’s own electromagnetic fields off the conducting surfaces, we generate intense transverse self-forces that pinch the beam toward a sharp focus~\cite{Sampath2021}. Crucially, the effect is inherently scalable: starting from a smaller beam strengthens the self-fields and therefore enhances the focusing, in stark contrast to conventional schemes where tighter focusing becomes increasingly difficult and therefore not scalable. Furthermore, unlike conventional schemes limited by external field strengths, this mechanism can push the beam to extreme densities, ultimately limited by the breakdown of the reflected-field approximation at near-solid densities. This self-aligned, compact platform is expected to enable unprecedented beam densities, providing a straightforward route to intense gamma-ray emission, strong ultrarelativistic beam-plasma interactions, enabling access to strong-field QED regimes~\cite{Sampath2021,MatheronCommPhys,YakimenkoPRL,TamburiniPRD} and ultimately pave the way to ultra compact focusing systems for future colliders.

To start with, let us consider a beam of rms length $\sigma_z$ and rms transverse size $\sigma_r$ as measured at the foil position. In the radiative regime ($\sigma_r\gtrsim\sigma_z$), the boundary-induced surface fields nearly double the azimuthal magnetic field while suppressing the radial electric field, yielding a strong net focusing force~\cite{corde2020HAL,Sampath2021,MatheronCommPhys} as a result of significant coherent transition radiation (CTR) at the surface. Near the beam axis, this interaction can be modeled as a thin lens with angular kick $\Delta x'=-x/f$ and effective focal length for a single foil
$f \approx 8\pi\varepsilon_0 \sigma_r^2 \mathcal{E}/e Q$~\cite{corde2020HAL} where $e$ is the elementary charge, $\varepsilon_0$ the vacuum permittivity, $Q$ and $\mathcal{E}$ the beam charge and particle energy. In the waist-at-foil configuration, for a multifoil target composed of $N$ foils, the beam divergence before the multifoil $\sigma_{r,0}'$ is increased after the multifoil to
\[\sigma_{r}'\simeq \sqrt{\sigma_{r,0}'^2 + \sigma_{r,0}^2\frac{N^2 \eta^2}{f^2}}\]
where $\sigma_{r,0}$ is the initial transverse beam size, and $\eta$ is the reformation factor described in Methods. The divergence increase occurring alongside the focusing effect motivates the use of post-target divergence as a sensitive observable for characterizing the focusing strength~\cite{corde2020HAL}.

The experiment was performed at SLAC's FACET-II facility~\cite{Yakimenko2019FACETII,Storey2024} [Fig.~\ref{FIGURE_1}(a)], where the multifoil target was placed at the end of the accelerator beamline. A set of final-focus quadrupoles concentrated the 10 GeV, 1 nC electron beam either on or before the target. Downstream quadrupoles were configured to re-image the beam in a parallel-to-point configuration, enabling precise measurement of the divergence. A dipole spectrometer dispersed the beam in energy onto a scintillator screen imaged by a camera, providing simultaneous diagnostics of beam divergence (horizontal axis) and energy (vertical axis).

When the beam strikes a conducting foil, the beam self-fields are reflected by the conduction electrons via coherent transition radiation~\cite{GinzburgTsytovich1990, jackson1998, Verzilov2000, Sampath2021}. This reflection cancels the naturally defocusing transverse electric field while reinforcing the azimuthal magnetic field, resulting in a net inward Lorentz force~\cite{corde2020HAL}. Between foils, the beam self-fields reform~\cite{Carron2000}, allowing each subsequent interface to act on a regenerated field. Stacking multiple thin foils amplifies the effect: each reflection further focuses the beam, creating a net cumulative focusing that can, in principle, reach extreme densities, greater than solid densities~\cite{Sampath2021}. Upon exiting the target and traveling past its best-focus position, the beam emerges from a sharp waist and expands rapidly [Fig.~\ref{FIGURE_1}(b)]. The increased post-focus divergence of the beam is reproduced in our fully kinetic particle-in-cell (PIC) simulations and serves as clear, highly sensitive observable for accurately characterizing the focusing strength.

Leveraging the tunable beam parameters at FACET-II, this study is organized as follows. (i) We first investigate elongated, uncompressed ``pencil'' beams traversing a 40-foil target; by varying the vacuum waist position, we observe a moderate divergence increase that reveals the onset of the self-focusing effect. (ii) We then examine longitudinally compressed beams where the focusing intensifies dramatically, marking the transition from a nonradiative to a radiative regime; these results show good agreement with PIC simulations as the number of foils is varied. (iii) Finally, we analyze a configuration where the beam waist is located upstream of the target, so that the initially expanding beam is recollimated by the multifoil, resulting in a clear reduction in divergence. In this case, the beam also carries an energy chirp, which provides additional insight into the longitudinal variation of the focusing force within the beam.

 \begin{figure}[t!] 
  \centering
  \includegraphics[width=7cm]{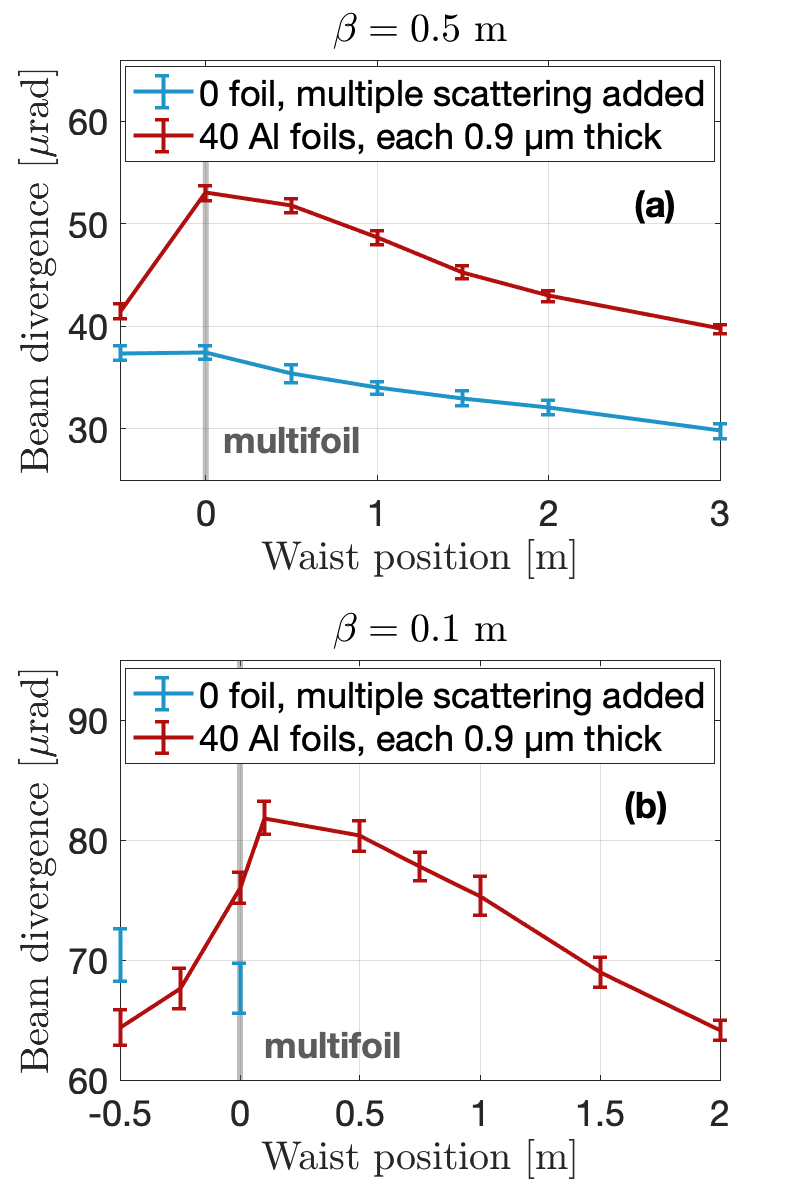}
  \caption{\justifying \textbf{Waist scan of the long-bunch regime around the multifoil position.} Horizontal divergence (rms) versus the waist position relative to the multifoil for $\beta=\SI{50}{cm}$ (a) and $\SI{10}{cm}$ (b). Red: 40-foil target. Blue: no-foil reference including the expected multiple-scattering contribution for 40 foils.}
  \label{FIGURE_2}
\end{figure}

Figure \ref{FIGURE_2} shows the results obtained in configuration (i). The uncompressed, pencil-like beam [$\sigma_r\approx\SI{25}{\micro\meter}$, $\sigma_z= \mathcal{O}(\SI{100}{\micro\meter})$], was sent onto a multifoil target composed of 40 ultra-thin aluminum foils, each \SI{0.9}{\micro\meter} thick and separated by \SI{100}{\micro\meter} using steel grids [Fig.~\ref{FIGURE_1}(c)]. The foil thickness was minimized to reduce scattering while remaining mechanically robust, and the spacing was chosen to allow beam self-fields to regenerate sufficiently between consecutive foils. Because repeated shots could progressively damage the foils, the target was mounted on a grid-shaped holder that enabled irradiation of fresh foil locations when needed. The beam waist position $z_\mathrm{w}$ was scanned so that the beam arrived diverging, collimated, or converging on the multifoil entrance, located at $z_\mathrm{MF}=0$. Figures~\ref{FIGURE_2}(a,b) show the measured rms horizontal divergence after the 40-foil target (red) as a function of waist position, for beams with $\beta$-functions of 50 cm and 10 cm (a Twiss parameter characterizing the envelope size of the beam~\cite{CourantSnyder}), together with the reference case without foils (blue; including the predicted multiple-scattering contribution for the 40 foils, see Methods).

When the beam waist position $z_{\rm w}$ is set to the multifoil position ($z_{\rm w} = 0$), its divergence increases because the beam is refocused to a new, narrower waist just downstream of the target before expanding rapidly, as illustrated in Fig.~\ref{FIGURE_1}(d). The effect is asymmetric with respect to the waist position. When the waist is located upstream, the beam arrives diverging and can be partially re-collimated by the multifoil, reducing the divergence [Fig.~\ref{FIGURE_1}(e)], as observed in Fig.~\ref{FIGURE_2}(b) for $z_\mathrm{w}\approx-\SI{0.5}{m}$. Conversely, when the waist is placed downstream, the beam size at the target is larger and self-fields are weaker, so the effect is reduced. In the long-bunch regime ($\sigma_z \gg \sigma_r$), evanescent spectral components dominate the foil-induced fields and the CTR emission is reduced. These fields remain confined to a narrow region ($\ll \sigma_z$) near the boundary and possess negligible radiative components~\cite{Sampath2021}, resulting in a moderate focusing kick only~\cite{Adler1982,Humphries1983,Humphries1988,Humphries1989,Fernsler1990}. To demonstrate the full potential of the scheme, the beam was compressed into a ``pancake'' shape  ($\sigma_z < \sigma_r$) with $\sigma_r\approx \SI{30}{\micro\meter}$ and $\sigma_z\simeq \SI{15}{\micro\meter}$, thereby increasing both the beam density and the strength of its self-fields. The multifoil then responds in the near-field coherent transition radiation (NF-CTR) regime, acting as an efficient mirror: the induced electromagnetic fields propagate away from the surface with an amplitude nearly equal to the beam self-fields, thereby dramatically enhancing the focusing effect~\cite{Sampath2021,corde2020HAL}. 

\begin{figure*}[t]
  \centering
  \includegraphics[width=18cm]{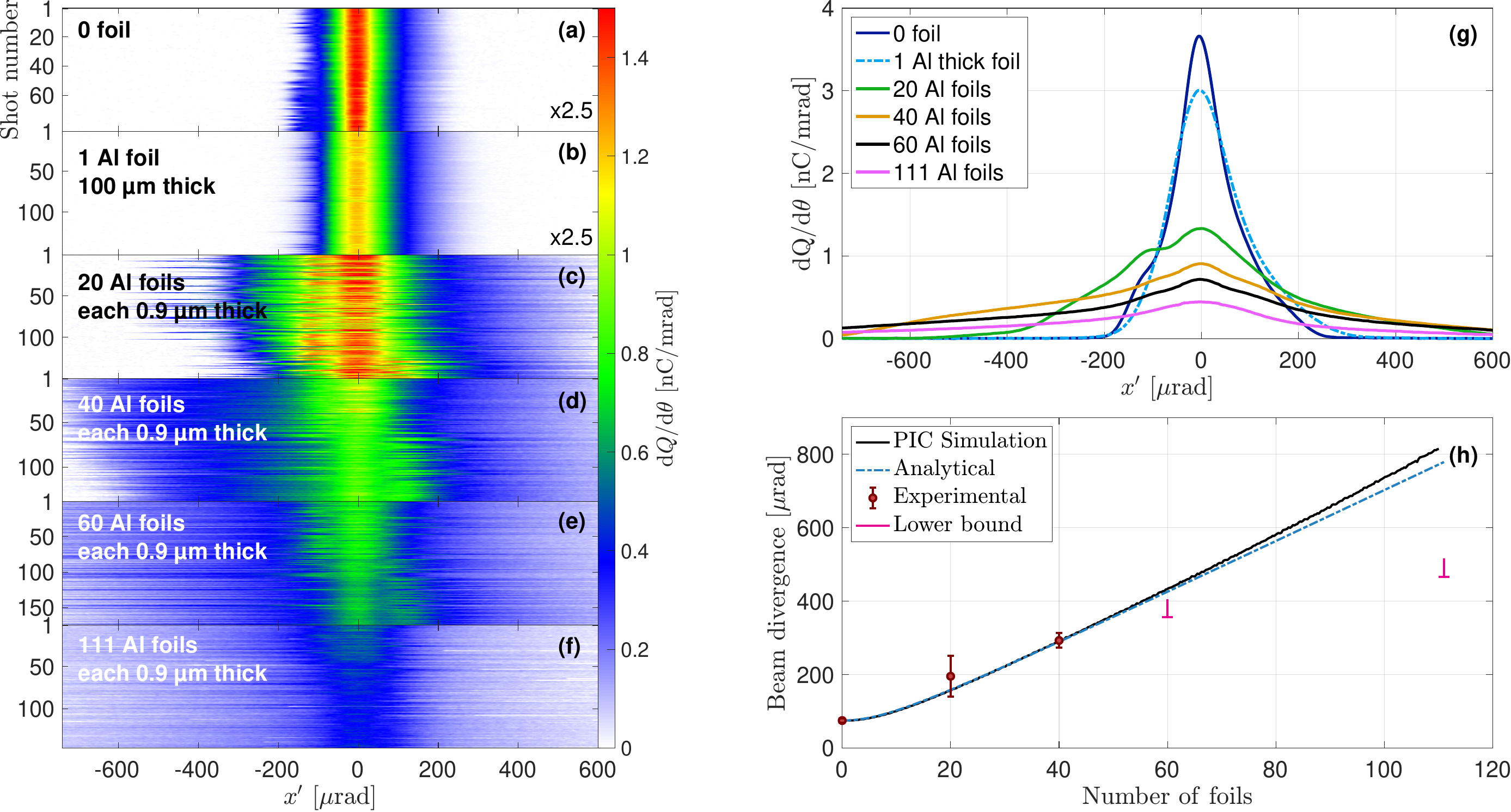}
  \caption{\justifying \textbf{Multifoil focusing of the compressed beam for varying foil numbers.} (a-f) Waterfall plots of the angular charge distribution for (a) 0 foil, (b) 1 thick (\SI{100}{\micro\meter}) aluminum foil, (c) 20, (d) 40, (e) 60, and (f) 111 thin aluminum foils. All foils are \SI{0.9}{\micro\meter} thick and are separated by \SI{100}{\micro\meter}. (g) Shot-averaged angular distributions corresponding to (a)-(f). (h) Beam divergence as a function of foil number. Red data points are the experimental measurements (with the error bar representing the standard deviation), pink data points indicate lower-bound experimental divergence values, the blue curve represents the analytical model and the black curve represents the PIC simulation.}
  \label{FIGURE_3}
\end{figure*}

Figure~\ref{FIGURE_3} summarizes results from 843 shots taken on four multifoil targets with different numbers of foils, together with a reference case without foils. In all configurations, the beam waist was positioned at the multifoil target [situation of Fig.~\ref{FIGURE_1}(d)]. Divergence is measured horizontally and energy vertically; the vertical divergence is not resolved. Figures~\ref{FIGURE_3}(a)--\ref{FIGURE_3}(f) show waterfall plots of the measured angular charge distributions $dQ/d\theta$ (integrated over energy), Fig.~\ref{FIGURE_3}(g) shows the corresponding shot-averaged profiles, and Fig.~\ref{FIGURE_3}(h) presents the rms divergence versus foil number.

As the number of foils $N$ increases, the beam divergence rises sharply, reflecting the cumulative nature of the focusing effect. Compared with the uncompressed case (shown in Fig.~\ref{FIGURE_2}), the divergence after the 40-foil target is roughly six times larger, demonstrating much stronger focusing. For large foil numbers ($N>10$), the divergence grows almost linearly with $N$. For $N=60$ and $N=111$, the focusing becomes so intense that the beam diverges beyond the collection aperture of the detector, causing the measured rms divergence to underestimate the true value. These data are therefore shown as lower bounds. Nevertheless, the continuous reduction in on-axis charge density with increasing foil number [Fig.~\ref{FIGURE_3}(g)] confirms that the focusing strength does not saturate, and that an increasing fraction of the beam is sent into the outer wings of the distribution. A good agreement is found with both PIC simulations and the analytical radiative model. The simulations and the radiative model assume a Gaussian beam prior to interaction with the multifoil target, with $\sigma_z=\SI{15}{\micro\meter}$ as measured by the X-band transverse deflecting cavity (X-TCAV) diagnostic (see Methods), and $\sigma_r=\SI{35}{\micro\meter}$, consistent with the wirescan measurements ($\sigma_r\approx \SI{30}{\micro\meter}$). The slightly larger transverse size used in the model can be explained by the non-Gaussian character of the beam, which can exhibit a narrower core accompanied by low-density tails. Moreover, the analytical model is radiative, and its agreement with the data confirms the radiative nature of the effect.

Multiple Coulomb scattering inside the foils can also increase divergence~\cite{Bethe1934,jackson1998}, but it is negligible for \SI{0.9}{\micro\meter} aluminum at \SI{10}{GeV} compared with the observed broadening (see Methods). This was verified by inserting a single \SI{100}{\micro\meter} thick foil---equivalent in total thickness to the 111-foil target---which produced only a minor divergence increase [Figs.~\ref{FIGURE_3}(b),\ref{FIGURE_3}(g)] . The dominant contribution in Fig.~\ref{FIGURE_3} that causes a divergence increase is therefore the multi-surface NF-CTR focusing mechanism. Collimation by the grid spacers is also excluded because the hole diameter of 1 mm corresponds to $\gg 10\sigma_r$, which produces no measurable divergence change for the beam configurations studied here.

\begin{figure*}[t!] 
  \centering
  \includegraphics[width=18cm]{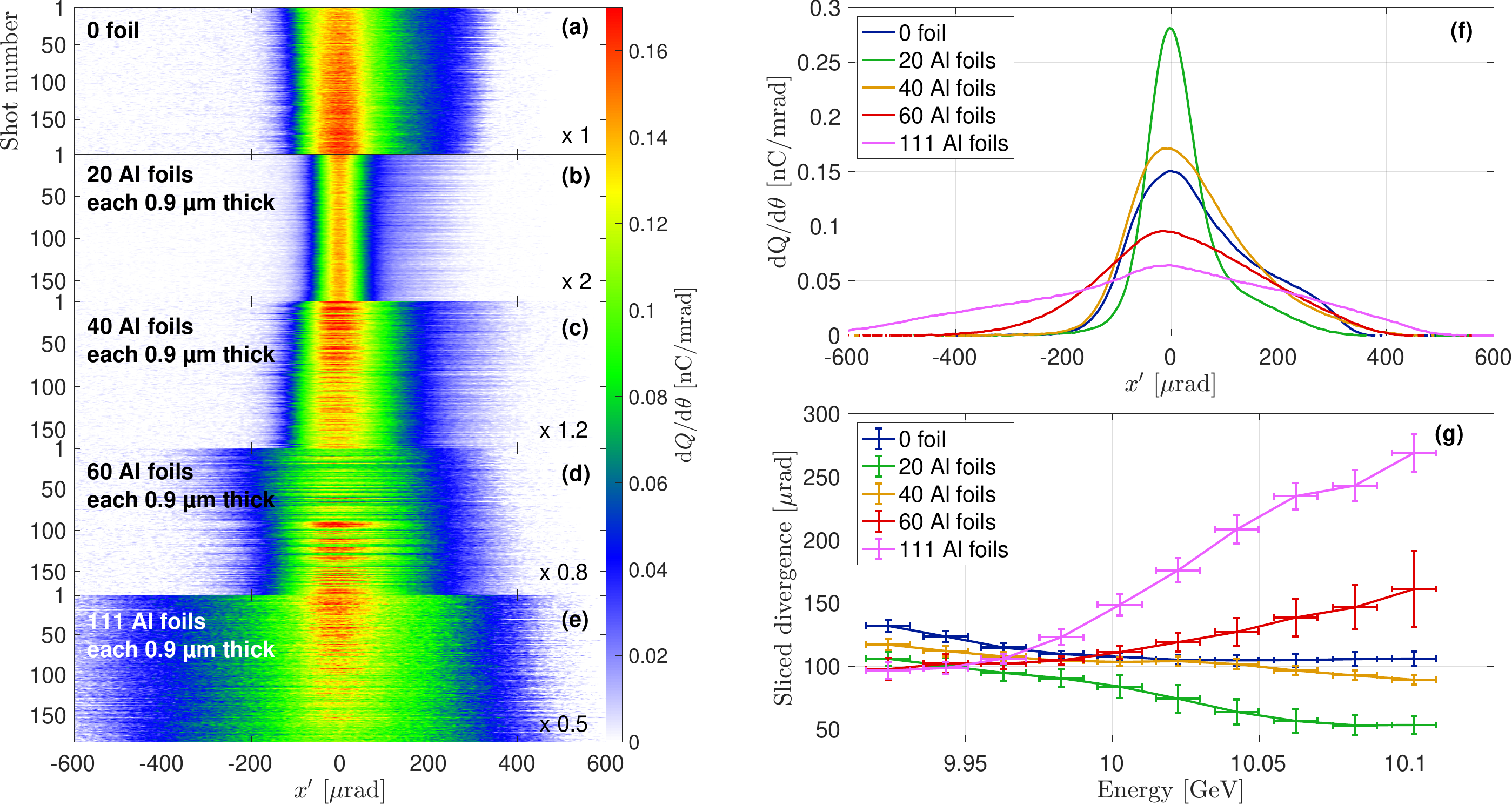} 
    \caption{\justifying \textbf{Multifoil focusing with beam waist upstream.} (a)-(e) Waterfall plots of the angular charge distribution of the \SI{10.06}{GeV} slice for (a) 0 foil, (b) 20, (c) 40, (d) 60, and (e) 111 thin aluminum foils. (f) Shot-averaged angular distributions corresponding to (a)-(e). (g) Divergence as a function of electron energy (increasing from the front to the back of the beam).}
  \label{FIGURE_4}
\end{figure*}

In the final configuration, the beam remained compressed but its waist was positioned \SI{50}{cm} upstream of the multifoil so that it arrived diverging on the target [Fig.~\ref{FIGURE_1}(e)]. Figure~\ref{FIGURE_4} summarizes results from 902 shots taken on four multifoil targets with different numbers of foils. Figures~\ref{FIGURE_4}(a)--\ref{FIGURE_4}(e) show the angular charge distributions at a single energy slice (\SI{10.0625}{GeV}$\pm$\SI{10}{MeV}) in waterfall form, while Fig.~\ref{FIGURE_4}(f) shows the averaged angular charge profiles. A comparison between the no-foil case [Fig.~\ref{FIGURE_4}(a)] and the 20-foil case [Fig.~\ref{FIGURE_4}(b)] reveals a clear reduction in divergence as expected, consistent with re-collimation of the beam by the multifoil target. This observation provides direct experimental evidence of the focusing dynamics in beam–multifoil interactions and rules out any alternative mechanism that could modify the divergence without producing focusing. As the number of foils increases further, the divergence rises again: stronger focusing overcompensates the initial divergence, moving the effective waist closer to the target and producing larger post-waist divergence. This optics-like behavior demonstrates that the effective focal length can be tuned by varying the number of foils, enabling either re-collimation or controlled post-target focusing. This is also evident in Fig.~\ref{FIGURE_4}(f), where the initially broad angular distribution becomes tightly concentrated around the axis for the 20-foil target. The effect, already hinted at for the uncompressed beam in Fig.~\ref{FIGURE_2}(b), becomes much stronger here, confirming the ability of the multifoil to re-collimate diverging beams—a key capability for beam guiding. 

Further insight is obtained from the same dataset, in which the beam was slightly over-compressed and exhibited an energy chirp, with low-energy electrons at the front and high-energy electrons at the back (see Methods and Extended Data Fig.~\ref{FIGURE_SM1}). Because the electron energy correlates with longitudinal position, we resolved the focusing dynamics along the beam by measuring angular charge distributions as a function of energy, as shown in Fig.~\ref{FIGURE_4}(g). A clear longitudinal dependence of the focusing strength is observed. In accordance with PIC simulations, electrons at the front of the bunch undergo minimal focusing because they experience relatively weak, short-duration reflected fields. By contrast, those at the back encounter fields generated by all preceding electrons, resulting in stronger focusing. This explains the evolution in Fig.~\ref{FIGURE_4}(g): at low energy (front of the beam), divergence decreases gradually with foil number, whereas at high energy (back of the beam), divergence first decreases strongly for a moderate foil number (e.g., 20) and then increases as focusing becomes overcompensated at high foil numbers (e.g., 111). This reproducible behavior supports operation in the NF-CTR regime, where the induced fields propagate backward and strengthen toward the bunch rear. In sharp contrast, in an evanescent-dominated non-radiative regime one would instead expect the strongest focusing near the bunch center.

To conclude, we have presented the first direct experimental evidence of multifoil-induced focusing of an ultrarelativistic, high-charge electron beam. The stability of FACET-II and the intrinsic auto-alignment of the mechanism--since by construction, the focusing fields are generated by and aligned with the beam itself--enable systematic exploration over a broad parameter range. Moderate focusing is observed for long, pencil-like beams in the non-radiative regime, while compressed pancake beams exhibit much stronger focusing enabled by NF-CTR. By shifting the waist upstream, it is further shown that an initially diverging beam can be recollimated by the multifoil focusing power. The good agreement with both PIC simulations and analytical models confirms that the underlying physics is well understood and controllable. The PIC simulation shown in Fig.~\ref{FIGURE_3}(h) for the 111-foil target illustrates the strength of the effect: the transverse size of the most focused beam slice is reduced by a factor of 11, from $\sigma_r=\SI{35}{\micro\meter}$ before the multifoil to $\sigma_r\simeq\SI{3}{\micro\meter}$ after it. The density increases by a factor of 120, from $n_e=2.2\times10^{16}~\mathrm{cm^{-3}}$ to $2.7\times10^{18}~\mathrm{cm^{-3}}$, and the effective focal length of the multifoil is $f=2.4~\mathrm{cm}$. Unlike conventional schemes limited by externally applied fields, this mechanism can enter a positive feedback loop regime. Focusing at each foil reduces the beam size before it reaches the next one, thereby increasing its density and self-fields as predicted by PIC simulations. The enhanced self-fields then strengthen the subsequent focusing, leading to a focusing positive feedback loop across the multifoil. This effect also ensures that all slices reach optimal focus despite the longitudinal dependence of the focusing effect. This regime could be reached through increased interfoil-spacing or higher beam densities.

The simplicity, robustness and versatility of this scheme for focusing, shaping and guiding high-energy particle beams makes it highly promising for frontier experiments in extreme plasma physics, relativistic laboratory astrophysics and strong-field QED. By coupling it with conventional focusing methods, plasma lenses~\cite{Doss2019} and beam compression~\cite{Emma2021}, one can envision pushing beam densities to solid-like levels--a territory previously considered out of experimental reach, where self-fields are strong enough to trigger laserless strong-field QED processes and efficiently convert beam energy into ultraintense gamma-ray bursts~\cite{Sampath2021, MatheronCommPhys}. Although our proof-of-principle experiment was performed with electrons, the underlying physics is universal and directly extendable to any high-current, compressed charged particle beams.

\bigskip
\noindent
{\textbf{Methods}}

{\small

\noindent
\textbf{Beamline, target and diagnostics} The FACET-II accelerator facility at SLAC National Accelerator Laboratory delivers \SI{10}{GeV} \SI{1}{nC} electron beams at a repetition rate of 10 Hz, focused by final-focus quadrupoles to beam sizes of $\sigma_r\approx\SI{25}{\micro\m}$ [Fig.~\ref{FIGURE_2}(a) data] and $\sigma_r\approx\SI{35}{\micro\m}$ (Figs.~\ref{FIGURE_3}-\ref{FIGURE_4} data). The waist position and $\beta$-function are controlled by the final-focus system. In the pencil low-energy-spread beam configuration of Fig.~\ref{FIGURE_2}, the beam is not compressed and $\sigma_z=\mathcal{O}(\SI{100}{\micro\m})$, while in the compressed beam configuration of Figs.~\ref{FIGURE_3}-\ref{FIGURE_4}, $\sigma_z\simeq\SI{15}{\micro\m}$ with a slight overcompression (see Methods on beam chirp measurement). The initial vacuum divergence ranges from $\approx\SI{35}{\micro rad}$ in Fig.~\ref{FIGURE_2}(a) to $\approx\SI{100}{\micro rad}$ in Fig.~\ref{FIGURE_4}. 

The multifoil target is made of \SI{0.9}{\micro\meter}-thick aluminum foils separated by \SI{100}{\micro\meter}-thick grid spacers made in stainless steel. The grid spacers have a pattern of holes, and when Al foils are damaged in one spot, the multifoil target is moved to a new hole position. A single \SI{100}{\micro\meter}-thick aluminum foil and four multifoil targets, of 20, 40, 60 and 111 foils, were available on the same target mover. 

The electron beam exiting the multifoil is imaged onto the detector using three quadrupoles located approximately \SI{3.7}{\meter}, \SI{5.9}{\meter} and \SI{8.1}{\meter} from the target, configured in a parallel-to-point optics. The corresponding transfer matrix satisfies $M_{11}=0$ and $M_{12}=15$ m, such that the horizontal rms beam size on the detector is $\sigma_x^{\mathrm{det}} = M_{12}\sigma_x'$, where $\sigma_x'$ denotes the horizontal divergence at the multifoil exit. Downstream, a dipole magnet located 13 m from the multifoil provides a nominal vertical dispersion of 60 mm at 10 GeV onto the detector, mapping the particle energy onto the vertical axis of the $\SI{50}{\micro\meter}$-thick YAG:Ce detector screen positioned 23 m from the multifoil. The resulting detector image thus encodes beam divergence horizontally and energy vertically, and enables the measurement of energy-resolved slice divergences. The vertical divergence is not measured, as the quadrupoles are configured for point-to-point imaging in the vertical direction ($M_{34}=0$) to optimize energy resolution.

\noindent
\textbf{Beam divergence calculation}
The beam rms divergence is extracted from the detector images using a standard image-processing pipeline. The background is subtracted from each image to ensure that the charge distribution goes to zero in absence of signal, and a 4-by-4-pixel median filter is applied to suppress hot pixels and gamma-induced noise. One-dimensional angular profiles are then obtained by integrating over energy (Figs.~\ref{FIGURE_2}-\ref{FIGURE_3}) or by selecting an energy slice (Fig.~\ref{FIGURE_4}). Residual noise is removed by applying a threshold cut at 2\% (Figs.~\ref{FIGURE_2}-\ref{FIGURE_3}) or 10\% (Fig.~\ref{FIGURE_4}) of the profile maximum. For Fig.~\ref{FIGURE_3}, only shots for which the uncalibrated sector 14 bunch length monitor value $B$ is between 3700 and 5300 (see Methods on Beam chirp measurement) and for which the beam energy in sectors 14 and 20 are within standard ranges are retained to remove outliers and shots during which the machine was not in standard operating conditions. For waterfall plots, individual shots are re-centered such that their barycenter satisfies $x' = 0$. The rms divergences are computed from the resulting angular profiles of individual shots.

For the strongest-focusing cases in Fig.~\ref{FIGURE_3} (60- and 111-foil configurations), the beam divergence exceeds the detector acceptance, leading to charge loss at the detector edges and an underestimation of the rms divergence. A conservative lower bound is therefore estimated by redistributing the missing charge into two narrow peaks located at $x'=\pm(\text{detector angular\ acceptance})/2$, such that the total charge is conserved. The corresponding rms value defines the lower-bound divergence reported in Fig.~\ref{FIGURE_3}(h).

\noindent
\textbf{Particle in cell simulation} 
The PIC simulation shown in Fig.~\ref{FIGURE_3} was performed using FBPIC~\cite{Lehe2016}. The initial electron-beam parameters were taken to be close to the experimental measurements: 10~GeV particle energy, 1~nC total charge, $\sigma_z = 15\,\rm \mu m$ rms length, $\sigma'_{x,0} =74.6\,\rm mrad$ initial divergence (as measured and assuming $\sigma'_{x,0} = \sigma'_{y,0}$). The transverse beam size $\sigma_r=35\ \mu$m was chosen to reproduce the experimentally observed focusing strength while being within reasonable experimental constraints. The beam is modeled with a Gaussian distribution using $5\times10^6$ macroparticles.

The multifoil target is modeled as a sequence of $0.9\ \mu$m-thick aluminum foils, each represented as a plasma with electron density $1.8\times10^{23}\ \mathrm{cm^{-3}}$, with 1 macroparticle per cell along the longitudinal ($z$) and azimuthal ($\theta$) directions and 2 macroparticles per cell along the transverse ($r$) direction. The foils are separated by $100\ \mu$m. The moving window of the simulation extends over $14\ \sigma_z$ longitudinally and $6\ \sigma_r$ transversely, with 16 916 cells in $z$, 500 cells in $r$, and a single azimuthal mode. Multiple Coulomb scattering and collisional effects are not included in the simulation. Due to the limited inter-foil spacing, the beam self-fields do not fully recover between successive foils, resulting in a reduced focusing strength. This effect is quantified by the reformation factor~\cite{MatheronThese}
\[\eta = \frac{B_\theta^{\mathrm{tot}}(\xi=0,r=1.6\sigma_r)_{\mathrm{subsequent\ foils}}}{B_\theta^{\mathrm{tot}}(\xi=0,r=1.6\sigma_r)_{\mathrm{first\ foil}}}\]
for which the simulation yields $\eta \simeq 0.51$ and stays almost constant at each foil surface. The electron beam at the exit of the multifoil PIC simulation is subsequently propagated in vacuum using a separate code until it reaches its post-target waist, allowing to calculate slice beam sizes, taking the rms of the profile with a cut at \SI{10}{\%} of the profile maximum, and allowing to estimate the maximum beam density reached at the post-target waist.

\noindent
\textbf{Multiple Coulomb scattering} 
Multiple Coulomb scattering (MCS) arises from successive small-angle deflections of electrons as they traverse matter (here, the aluminum foils). It is well described statistically by a Gaussian angular distribution with zero mean and rms width
\begin{equation}
\theta_\mathrm{MCS}=\frac{13.6}{\mathcal{E}\,[\mathrm{MeV}]}\sqrt{\frac{d}{X_0}}\left(1+0.038\ln\frac{d}{X_0}\right),
\end{equation}
where $d$ is the material thickness and $X_0$ its radiation length ($X_0=\SI{8.9}{cm}$ for aluminum). In contrast to the NF--CTR focusing, which is a surface effect, MCS develops in the bulk of the material. The multifoil target is therefore designed to maximize the number of surfaces while minimizing the total amount of material traversed by the beam. Using $\SI{0.9}{\micro\meter}$-thick foils keeps MCS small while remaining sufficiently robust to be handled and assembled into the multifoil structure. For a single foil, $\theta_\mathrm{MCS}\simeq \SI{2.4}{\micro rad}$. In the most extreme configuration (111 foils), the accumulated scattering reaches only $\theta_\mathrm{MCS}\simeq\SI{34}{\micro rad}$. Adding this contribution in quadrature to the initial vacuum divergence $\sigma_{r,0}' = \SI{74.6}{\micro rad}$ yields an expected divergence of $\sqrt{\sigma_{r,0}'^2 + \theta_\mathrm{MCS}^2}\simeq \SI{82}{\micro rad}$, a negligible effect compared to the experimentally measured divergence of at least \SI{480}{\micro rad} (lower bound) or the simulated divergence of $\approx\SI{800}{\micro rad}$ in PIC for 111 foils. The no-foil reference data in Fig.~\ref{FIGURE_2} corresponds to the experimentally measured divergence in vacuum, to which is added in quadrature the MCS effect of 40 Al foils.

\noindent
\textbf{Beam chirp measurement} 
The longitudinal phase space (LPS) of the beam in experiments at FACET-II correlates very closely with the uncalibrated value, defined as $B$ here, measured by the bunch length monitor located in sector 14 at the end of the second bunch compressor of the accelerator. This monitor consists of an infrared pyrometer that measures the total yield of coherent edge radiation emitted by the beam when exiting the last bend magnet of the chicane~\cite{Loos2007}. The resultant signal is directly correlated with the bunch length at the location of the bunch length monitor, a shorter bunch generating a larger signal. For the data of Fig.~\ref{FIGURE_3}, $\langle B\rangle=4431$ and for the data of Fig.~\ref{FIGURE_4},  $\langle B\rangle=3739$, with a typical standard deviation in $B$ of about 500 in each dataset. The energy chirp of the beam was measured in the range from $B=3100$ to $B=5100$, and the measurement showed that in this range the beam is slightly over-compressed by the final bunch compressor in sector 20, with lower energies at the front of the bunch and higher energies at the back at the interaction point (IP). The LPS core follows a linear relationship [see black line in Extended Data Fig.~\ref{FIGURE_SM1}(b)] of the form $z=c(\mathcal{E}-\mathcal{E}_0)$, with $\mathcal{E}$ the particle energy and $\mathcal{E}_0=\SI{10}{GeV}$ the central beam energy, with the energy chirp $c$ ranging from \SI{-0.25}{\micro\m\per\MeV} to \SI{-0.45}{\micro\m\per\MeV} [see Extended Data Fig.~\ref{FIGURE_SM1}(a)]. For $B=4500$, relevant for the data of Fig.~\ref{FIGURE_3}, the experimental reconstructed LPS and current profile are shown in Extended Data Figs.~\ref{FIGURE_SM1}(b)-(c), with an rms bunch length of \SI{16}{\micro\m} and an energy chirp of \SI{-0.39}{\micro\m\per\MeV}.

The energy chirp and bunch length at the IP were extracted from X-band transverse deflecting cavity measurements. The X-TCAV is a device placed upstream of the FACET experimental area, providing a time-dependent transverse kick to the electron beam via an alternating transverse electric field in the X-band frequency range. When the zero crossing of this field is synchronized with the time of arrival of the electron bunch center, each longitudinal position of the particle bunch is mapped to a different horizontal position on the detector screen, the same detector as used for multifoil divergence measurements. This detector screen (electron diagnostic camera in Fig.~\ref{FIGURE_1}) now encodes longitudinal position or time along the horizontal axis when X-TCAV is on, and particle energy along the vertical axis because of the dispersion provided by the spectrometer dipole (also shown in Fig.~\ref{FIGURE_1}), thereby revealing the LPS of the bunch. Both zero crossings are measured separately, and the rms of the projection on the horizontal axis is corrected for transverse contributions by subtracting the X-TCAV off measurement in quadrature, an operation allowing to deconvolve the horizontal beam size that can limit the X-TCAV resolution. A complete LPS reconstruction [Extended Data Fig.~\ref{FIGURE_SM1}(b)] was carried out by analyzing independently different energy bands about 17 MeV in width, each energy band being modeled by a Gaussian profile with the rms bunch length and centroid longitudinal position obtained from the X-TCAV deconvolution. This method also corrects for $x-\mathcal{E}$ correlations that may exist at the detector screen with X-TCAV off, and that may distort the LPS measurement otherwise. The energy chirp of the beam is obtained from linear fits in the energy range from \SI{9.975}{GeV} to \SI{10.750}{GeV}.

\bigskip
\noindent
\textbf{Data availability}

{\small
\noindent
The data that support the findings of this study are available from AM and DS upon request.}

\bigskip
\noindent
\textbf{Code availability}

{\small
\noindent
The FBPIC code has been used to generate the PIC simulation presented in this manuscript. This code is available in open source~\cite{Lehe2016,FBPICGithub}.
}

\bigskip
\noindent
\textbf{Acknowledgments}

{\small
\noindent
The FACET-II E-332 beam-multifoil experiment is operated with funding from the U.S. Department of Energy under Contract No. DE-AC02-76SF00515. This work was supported at LOA by the ANR (g4QED project, Grant No. ANR-23-CE30-0011). AM was supported by the ANR under the program ``Investissements d'Avenir'' (Grant No. ANR-18-EURE-0014). AM and MG were supported by the France-Stanford Center for Interdisciplinary Studies for travels to SLAC National Accelerator Laboratory. This work was supported at UCLA by the U.S. Department of Energy through Grant No. DE-SC0010064.
}

\bigskip
\noindent
\textbf{Author contributions}

{\small
\noindent
All authors contributed to the work presented in this paper.
}

\bigskip
\noindent
\textbf{Competing interests}

\noindent
{\small
\noindent
The authors declare no competing interests.
}

\newpage

\begin{smfigure*}[t!] 
  \centering
  \includegraphics[width=18cm]{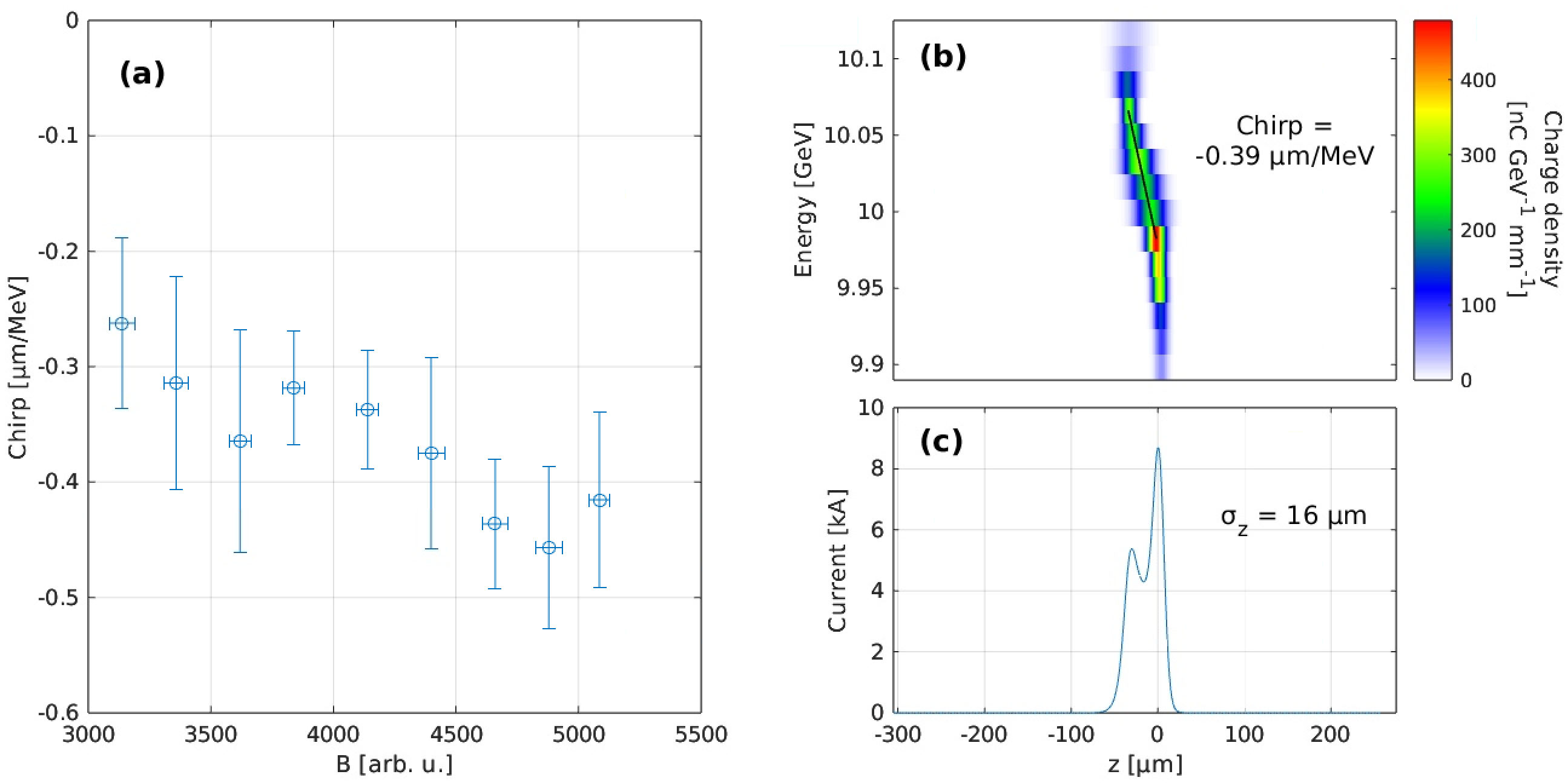} 
    \caption{\justifying \textbf{Beam chirp at different compressions and longitudinal phase space.} (a) Variation of energy chirp with the uncalibrated value $B$ of the sector 14 bunch length monitor. Longitudinal phase space (b) and current profile (c) reconstructed from X-TCAV measurements for $B = 4500$ (using all shots with $B$ in the range 4400 to 4600). Positive $z$ is the front of the beam. The black solid line in (b) is a linear fit with a slope of \SI{-0.39}{\micro\m\per\MeV}.}
  \label{FIGURE_SM1}
\end{smfigure*}


\begin{thebibliography}{34}%
\makeatletter
\providecommand \@ifxundefined [1]{%
 \@ifx{#1\undefined}
}%
\providecommand \@ifnum [1]{%
 \ifnum #1\expandafter \@firstoftwo
 \else \expandafter \@secondoftwo
 \fi
}%
\providecommand \@ifx [1]{%
 \ifx #1\expandafter \@firstoftwo
 \else \expandafter \@secondoftwo
 \fi
}%
\providecommand \natexlab [1]{#1}%
\providecommand \enquote  [1]{``#1''}%
\providecommand \bibnamefont  [1]{#1}%
\providecommand \bibfnamefont [1]{#1}%
\providecommand \citenamefont [1]{#1}%
\providecommand \href@noop [0]{\@secondoftwo}%
\providecommand \href [0]{\begingroup \@sanitize@url \@href}%
\providecommand \@href[1]{\@@startlink{#1}\@@href}%
\providecommand \@@href[1]{\endgroup#1\@@endlink}%
\providecommand \@sanitize@url [0]{\catcode `\\12\catcode `\$12\catcode `\&12\catcode `\#12\catcode `\^12\catcode `\_12\catcode `\%12\relax}%
\providecommand \@@startlink[1]{}%
\providecommand \@@endlink[0]{}%
\providecommand \url  [0]{\begingroup\@sanitize@url \@url }%
\providecommand \@url [1]{\endgroup\@href {#1}{\urlprefix }}%
\providecommand \urlprefix  [0]{URL }%
\providecommand \Eprint [0]{\href }%
\providecommand \doibase [0]{https://doi.org/}%
\providecommand \selectlanguage [0]{\@gobble}%
\providecommand \bibinfo  [0]{\@secondoftwo}%
\providecommand \bibfield  [0]{\@secondoftwo}%
\providecommand \translation [1]{[#1]}%
\providecommand \BibitemOpen [0]{}%
\providecommand \bibitemStop [0]{}%
\providecommand \bibitemNoStop [0]{.\EOS\space}%
\providecommand \EOS [0]{\spacefactor3000\relax}%
\providecommand \BibitemShut  [1]{\csname bibitem#1\endcsname}%
\let\auto@bib@innerbib\@empty
\bibitem [{\citenamefont {Bilderback}\ \emph {et~al.}(2005)\citenamefont {Bilderback}, \citenamefont {Elleaume},\ and\ \citenamefont {Weckert}}]{Bilderback2005}%
  \BibitemOpen
  \bibfield  {author} {\bibinfo {author} {\bibfnamefont {D.~H.}\ \bibnamefont {Bilderback}}, \bibinfo {author} {\bibfnamefont {P.}~\bibnamefont {Elleaume}},\ and\ \bibinfo {author} {\bibfnamefont {E.}~\bibnamefont {Weckert}},\ }\href {https://doi.org/10.1088/0953-4075/38/9/022} {\bibfield  {journal} {\bibinfo  {journal} {J. Phys. B: At. Mol. Opt. Phys.}\ }\textbf {\bibinfo {volume} {38}},\ \bibinfo {pages} {S773} (\bibinfo {year} {2005})}\BibitemShut {NoStop}%
\bibitem [{\citenamefont {Mourou}\ \emph {et~al.}(2006)\citenamefont {Mourou}, \citenamefont {Tajima},\ and\ \citenamefont {Bulanov}}]{MourouExtremeBeams2006}%
  \BibitemOpen
  \bibfield  {author} {\bibinfo {author} {\bibfnamefont {G.~A.}\ \bibnamefont {Mourou}}, \bibinfo {author} {\bibfnamefont {T.}~\bibnamefont {Tajima}},\ and\ \bibinfo {author} {\bibfnamefont {S.~V.}\ \bibnamefont {Bulanov}},\ }\href {https://doi.org/10.1103/RevModPhys.78.309} {\bibfield  {journal} {\bibinfo  {journal} {Rev. Mod. Phys.}\ }\textbf {\bibinfo {volume} {78}},\ \bibinfo {pages} {309} (\bibinfo {year} {2006})}\BibitemShut {NoStop}%
\bibitem [{\citenamefont {Marklund}\ and\ \citenamefont {Shukla}(2006)}]{Marklund2006}%
  \BibitemOpen
  \bibfield  {author} {\bibinfo {author} {\bibfnamefont {M.}~\bibnamefont {Marklund}}\ and\ \bibinfo {author} {\bibfnamefont {P.~K.}\ \bibnamefont {Shukla}},\ }\href {https://doi.org/10.1103/RevModPhys.78.591} {\bibfield  {journal} {\bibinfo  {journal} {Rev. Mod. Phys.}\ }\textbf {\bibinfo {volume} {78}},\ \bibinfo {pages} {591} (\bibinfo {year} {2006})}\BibitemShut {NoStop}%
\bibitem [{\citenamefont {Ruffini}\ \emph {et~al.}(2010)\citenamefont {Ruffini}, \citenamefont {Vereshchagin},\ and\ \citenamefont {Xue}}]{Ruffini2010}%
  \BibitemOpen
  \bibfield  {author} {\bibinfo {author} {\bibfnamefont {R.}~\bibnamefont {Ruffini}}, \bibinfo {author} {\bibfnamefont {G.}~\bibnamefont {Vereshchagin}},\ and\ \bibinfo {author} {\bibfnamefont {S.-S.}\ \bibnamefont {Xue}},\ }\href {https://doi.org/10.1016/j.physrep.2009.10.004} {\bibfield  {journal} {\bibinfo  {journal} {Phys. Rep.}\ }\textbf {\bibinfo {volume} {487}},\ \bibinfo {pages} {1} (\bibinfo {year} {2010})}\BibitemShut {NoStop}%
\bibitem [{\citenamefont {Di~Piazza}\ \emph {et~al.}(2012)\citenamefont {Di~Piazza}, \citenamefont {M{\"u}ller}, \citenamefont {Hatsagortsyan},\ and\ \citenamefont {Keitel}}]{DiPiazza2012}%
  \BibitemOpen
  \bibfield  {author} {\bibinfo {author} {\bibfnamefont {A.}~\bibnamefont {Di~Piazza}}, \bibinfo {author} {\bibfnamefont {C.}~\bibnamefont {M{\"u}ller}}, \bibinfo {author} {\bibfnamefont {K.~Z.}\ \bibnamefont {Hatsagortsyan}},\ and\ \bibinfo {author} {\bibfnamefont {C.~H.}\ \bibnamefont {Keitel}},\ }\href {https://doi.org/10.1103/RevModPhys.84.1177} {\bibfield  {journal} {\bibinfo  {journal} {Rev. Mod. Phys.}\ }\textbf {\bibinfo {volume} {84}},\ \bibinfo {pages} {1177} (\bibinfo {year} {2012})}\BibitemShut {NoStop}%
\bibitem [{\citenamefont {Corde}\ \emph {et~al.}(2013)\citenamefont {Corde}, \citenamefont {Ta~Phuoc}, \citenamefont {Lambert}, \citenamefont {Fitour}, \citenamefont {Malka}, \citenamefont {Rousse}, \citenamefont {Beck},\ and\ \citenamefont {Lefebvre}}]{Corde2013}%
  \BibitemOpen
  \bibfield  {author} {\bibinfo {author} {\bibfnamefont {S.}~\bibnamefont {Corde}}, \bibinfo {author} {\bibfnamefont {K.}~\bibnamefont {Ta~Phuoc}}, \bibinfo {author} {\bibfnamefont {G.}~\bibnamefont {Lambert}}, \bibinfo {author} {\bibfnamefont {R.}~\bibnamefont {Fitour}}, \bibinfo {author} {\bibfnamefont {V.}~\bibnamefont {Malka}}, \bibinfo {author} {\bibfnamefont {A.}~\bibnamefont {Rousse}}, \bibinfo {author} {\bibfnamefont {A.}~\bibnamefont {Beck}},\ and\ \bibinfo {author} {\bibfnamefont {E.}~\bibnamefont {Lefebvre}},\ }\href {https://doi.org/10.1103/RevModPhys.85.1} {\bibfield  {journal} {\bibinfo  {journal} {Rev. Mod. Phys.}\ }\textbf {\bibinfo {volume} {85}},\ \bibinfo {pages} {1} (\bibinfo {year} {2013})}\BibitemShut {NoStop}%
\bibitem [{\citenamefont {Ullrich}\ \emph {et~al.}(2012)\citenamefont {Ullrich}, \citenamefont {Rudenko},\ and\ \citenamefont {Moshammer}}]{Ullrich2012}%
  \BibitemOpen
  \bibfield  {author} {\bibinfo {author} {\bibfnamefont {J.}~\bibnamefont {Ullrich}}, \bibinfo {author} {\bibfnamefont {A.}~\bibnamefont {Rudenko}},\ and\ \bibinfo {author} {\bibfnamefont {R.}~\bibnamefont {Moshammer}},\ }\href {https://doi.org/10.1146/annurev-physchem-032511-143720} {\bibfield  {journal} {\bibinfo  {journal} {Annu. Rev. Phys. Chem.}\ }\textbf {\bibinfo {volume} {63}},\ \bibinfo {pages} {635} (\bibinfo {year} {2012})}\BibitemShut {NoStop}%
\bibitem [{\citenamefont {Uzdensky}\ \emph {et~al.}(2019)\citenamefont {Uzdensky}, \citenamefont {Begelman}, \citenamefont {Beloborodov}, \citenamefont {Blandford}, \citenamefont {Boldyrev}, \citenamefont {Cerutti}, \citenamefont {Fiuza}, \citenamefont {Giannios}, \citenamefont {Grismayer}, \citenamefont {Kunz}, \citenamefont {Loureiro}, \citenamefont {Lyutikov}, \citenamefont {Medvedev}, \citenamefont {Petropoulou}, \citenamefont {Philippov}, \citenamefont {Quataert}, \citenamefont {Schekochihin}, \citenamefont {Schoeffler}, \citenamefont {Silva}, \citenamefont {Sironi}, \citenamefont {Spitkovsky}, \citenamefont {Werner}, \citenamefont {Zhdankin}, \citenamefont {Zrake},\ and\ \citenamefont {Zweibel}}]{UzdenskiArXiv2019}%
  \BibitemOpen
  \bibfield  {author} {\bibinfo {author} {\bibfnamefont {D.}~\bibnamefont {Uzdensky}}, \bibinfo {author} {\bibfnamefont {M.}~\bibnamefont {Begelman}}, \bibinfo {author} {\bibfnamefont {A.}~\bibnamefont {Beloborodov}}, \bibinfo {author} {\bibfnamefont {R.}~\bibnamefont {Blandford}}, \bibinfo {author} {\bibfnamefont {S.}~\bibnamefont {Boldyrev}}, \bibinfo {author} {\bibfnamefont {B.}~\bibnamefont {Cerutti}}, \bibinfo {author} {\bibfnamefont {F.}~\bibnamefont {Fiuza}}, \bibinfo {author} {\bibfnamefont {D.}~\bibnamefont {Giannios}}, \bibinfo {author} {\bibfnamefont {T.}~\bibnamefont {Grismayer}}, \bibinfo {author} {\bibfnamefont {M.}~\bibnamefont {Kunz}}, \bibinfo {author} {\bibfnamefont {N.}~\bibnamefont {Loureiro}}, \bibinfo {author} {\bibfnamefont {M.}~\bibnamefont {Lyutikov}}, \bibinfo {author} {\bibfnamefont {M.}~\bibnamefont {Medvedev}}, \bibinfo {author} {\bibfnamefont {M.}~\bibnamefont {Petropoulou}}, \bibinfo {author} {\bibfnamefont {A.}~\bibnamefont {Philippov}}, \bibinfo {author} {\bibfnamefont
  {E.}~\bibnamefont {Quataert}}, \bibinfo {author} {\bibfnamefont {A.}~\bibnamefont {Schekochihin}}, \bibinfo {author} {\bibfnamefont {K.}~\bibnamefont {Schoeffler}}, \bibinfo {author} {\bibfnamefont {L.}~\bibnamefont {Silva}}, \bibinfo {author} {\bibfnamefont {L.}~\bibnamefont {Sironi}}, \bibinfo {author} {\bibfnamefont {A.}~\bibnamefont {Spitkovsky}}, \bibinfo {author} {\bibfnamefont {G.}~\bibnamefont {Werner}}, \bibinfo {author} {\bibfnamefont {V.}~\bibnamefont {Zhdankin}}, \bibinfo {author} {\bibfnamefont {J.}~\bibnamefont {Zrake}},\ and\ \bibinfo {author} {\bibfnamefont {E.}~\bibnamefont {Zweibel}},\ }\href {https://doi.org/10.48550/arXiv.1903.05328} {\bibinfo {title} {Extreme plasma astrophysics}},\ \bibinfo {howpublished} {arXiv:1903.05328} (\bibinfo {year} {2019})\BibitemShut {NoStop}%
\bibitem [{\citenamefont {Xu}\ \emph {et~al.}(2021)\citenamefont {Xu}, \citenamefont {Cesar}, \citenamefont {Corde}, \citenamefont {Yakimenko}, \citenamefont {Hogan}, \citenamefont {Joshi}, \citenamefont {Marinelli},\ and\ \citenamefont {Mori}}]{Xu2021}%
  \BibitemOpen
  \bibfield  {author} {\bibinfo {author} {\bibfnamefont {X.}~\bibnamefont {Xu}}, \bibinfo {author} {\bibfnamefont {D.~B.}\ \bibnamefont {Cesar}}, \bibinfo {author} {\bibfnamefont {S.}~\bibnamefont {Corde}}, \bibinfo {author} {\bibfnamefont {V.}~\bibnamefont {Yakimenko}}, \bibinfo {author} {\bibfnamefont {M.~J.}\ \bibnamefont {Hogan}}, \bibinfo {author} {\bibfnamefont {C.}~\bibnamefont {Joshi}}, \bibinfo {author} {\bibfnamefont {A.}~\bibnamefont {Marinelli}},\ and\ \bibinfo {author} {\bibfnamefont {W.~B.}\ \bibnamefont {Mori}},\ }\href {https://doi.org/10.1103/PhysRevLett.126.094801} {\bibfield  {journal} {\bibinfo  {journal} {Phys. Rev. Lett.}\ }\textbf {\bibinfo {volume} {126}},\ \bibinfo {pages} {094801} (\bibinfo {year} {2021})}\BibitemShut {NoStop}%
\bibitem [{\citenamefont {Telnov}(1990)}]{Telnov1990}%
  \BibitemOpen
  \bibfield  {author} {\bibinfo {author} {\bibfnamefont {V.~I.}\ \bibnamefont {Telnov}},\ }\href {https://doi.org/10.1016/0168-9002(90)91826-W} {\bibfield  {journal} {\bibinfo  {journal} {Nucl. Instrum. Methods Phys. Res. A}\ }\textbf {\bibinfo {volume} {294}},\ \bibinfo {pages} {72} (\bibinfo {year} {1990})}\BibitemShut {NoStop}%
\bibitem [{\citenamefont {Sampath}\ \emph {et~al.}(2021)\citenamefont {Sampath}, \citenamefont {Davoine}, \citenamefont {Corde}, \citenamefont {Gremillet}, \citenamefont {Gilljohann}, \citenamefont {Sangal}, \citenamefont {Keitel}, \citenamefont {Ariniello}, \citenamefont {Cary}, \citenamefont {Ekerfelt}, \citenamefont {Emma}, \citenamefont {Fiuza}, \citenamefont {Fujii}, \citenamefont {Hogan}, \citenamefont {Joshi}, \citenamefont {Knetsch}, \citenamefont {Kononenko}, \citenamefont {Lee}, \citenamefont {Litos}, \citenamefont {Marsh}, \citenamefont {Nie}, \citenamefont {O'Shea}, \citenamefont {Peterson}, \citenamefont {San Miguel~Claveria}, \citenamefont {Storey}, \citenamefont {Wu}, \citenamefont {Xu}, \citenamefont {Zhang},\ and\ \citenamefont {Tamburini}}]{Sampath2021}%
  \BibitemOpen
  \bibfield  {author} {\bibinfo {author} {\bibfnamefont {A.}~\bibnamefont {Sampath}}, \bibinfo {author} {\bibfnamefont {X.}~\bibnamefont {Davoine}}, \bibinfo {author} {\bibfnamefont {S.}~\bibnamefont {Corde}}, \bibinfo {author} {\bibfnamefont {L.}~\bibnamefont {Gremillet}}, \bibinfo {author} {\bibfnamefont {M.}~\bibnamefont {Gilljohann}}, \bibinfo {author} {\bibfnamefont {M.}~\bibnamefont {Sangal}}, \bibinfo {author} {\bibfnamefont {C.~H.}\ \bibnamefont {Keitel}}, \bibinfo {author} {\bibfnamefont {R.}~\bibnamefont {Ariniello}}, \bibinfo {author} {\bibfnamefont {J.}~\bibnamefont {Cary}}, \bibinfo {author} {\bibfnamefont {H.}~\bibnamefont {Ekerfelt}}, \bibinfo {author} {\bibfnamefont {C.}~\bibnamefont {Emma}}, \bibinfo {author} {\bibfnamefont {F.}~\bibnamefont {Fiuza}}, \bibinfo {author} {\bibfnamefont {H.}~\bibnamefont {Fujii}}, \bibinfo {author} {\bibfnamefont {M.}~\bibnamefont {Hogan}}, \bibinfo {author} {\bibfnamefont {C.}~\bibnamefont {Joshi}}, \bibinfo {author} {\bibfnamefont {A.}~\bibnamefont {Knetsch}},
  \bibinfo {author} {\bibfnamefont {O.}~\bibnamefont {Kononenko}}, \bibinfo {author} {\bibfnamefont {V.}~\bibnamefont {Lee}}, \bibinfo {author} {\bibfnamefont {M.}~\bibnamefont {Litos}}, \bibinfo {author} {\bibfnamefont {K.}~\bibnamefont {Marsh}}, \bibinfo {author} {\bibfnamefont {Z.}~\bibnamefont {Nie}}, \bibinfo {author} {\bibfnamefont {B.}~\bibnamefont {O'Shea}}, \bibinfo {author} {\bibfnamefont {J.~R.}\ \bibnamefont {Peterson}}, \bibinfo {author} {\bibfnamefont {P.}~\bibnamefont {San Miguel~Claveria}}, \bibinfo {author} {\bibfnamefont {D.}~\bibnamefont {Storey}}, \bibinfo {author} {\bibfnamefont {Y.}~\bibnamefont {Wu}}, \bibinfo {author} {\bibfnamefont {X.}~\bibnamefont {Xu}}, \bibinfo {author} {\bibfnamefont {C.}~\bibnamefont {Zhang}},\ and\ \bibinfo {author} {\bibfnamefont {M.}~\bibnamefont {Tamburini}},\ }\href {https://doi.org/10.1103/PhysRevLett.126.064801} {\bibfield  {journal} {\bibinfo  {journal} {Phys. Rev. Lett.}\ }\textbf {\bibinfo {volume} {126}},\ \bibinfo {pages} {064801} (\bibinfo {year}
  {2021})}\BibitemShut {NoStop}%
\bibitem [{\citenamefont {Matheron}\ \emph {et~al.}(2023)\citenamefont {Matheron}, \citenamefont {San Miguel~Claveria}, \citenamefont {Ariniello}, \citenamefont {Ekerfelt}, \citenamefont {Fiuza}, \citenamefont {Gessner}, \citenamefont {Gilljohann}, \citenamefont {Hogan}, \citenamefont {Keitel}, \citenamefont {Knetsch}, \citenamefont {Litos}, \citenamefont {Mankovska}, \citenamefont {Montefiori}, \citenamefont {Nie}, \citenamefont {O'Shea}, \citenamefont {Peterson}, \citenamefont {Storey}, \citenamefont {Wu}, \citenamefont {Xu}, \citenamefont {Zakharova}, \citenamefont {Davoine}, \citenamefont {Gremillet}, \citenamefont {Tamburini},\ and\ \citenamefont {Corde}}]{MatheronCommPhys}%
  \BibitemOpen
  \bibfield  {author} {\bibinfo {author} {\bibfnamefont {A.}~\bibnamefont {Matheron}}, \bibinfo {author} {\bibfnamefont {P.}~\bibnamefont {San Miguel~Claveria}}, \bibinfo {author} {\bibfnamefont {R.}~\bibnamefont {Ariniello}}, \bibinfo {author} {\bibfnamefont {H.}~\bibnamefont {Ekerfelt}}, \bibinfo {author} {\bibfnamefont {F.}~\bibnamefont {Fiuza}}, \bibinfo {author} {\bibfnamefont {S.}~\bibnamefont {Gessner}}, \bibinfo {author} {\bibfnamefont {M.~F.}\ \bibnamefont {Gilljohann}}, \bibinfo {author} {\bibfnamefont {M.~J.}\ \bibnamefont {Hogan}}, \bibinfo {author} {\bibfnamefont {C.~H.}\ \bibnamefont {Keitel}}, \bibinfo {author} {\bibfnamefont {A.}~\bibnamefont {Knetsch}}, \bibinfo {author} {\bibfnamefont {M.}~\bibnamefont {Litos}}, \bibinfo {author} {\bibfnamefont {Y.}~\bibnamefont {Mankovska}}, \bibinfo {author} {\bibfnamefont {S.}~\bibnamefont {Montefiori}}, \bibinfo {author} {\bibfnamefont {Z.}~\bibnamefont {Nie}}, \bibinfo {author} {\bibfnamefont {B.}~\bibnamefont {O'Shea}}, \bibinfo {author} {\bibfnamefont
  {J.~R.}\ \bibnamefont {Peterson}}, \bibinfo {author} {\bibfnamefont {D.}~\bibnamefont {Storey}}, \bibinfo {author} {\bibfnamefont {Y.}~\bibnamefont {Wu}}, \bibinfo {author} {\bibfnamefont {X.}~\bibnamefont {Xu}}, \bibinfo {author} {\bibfnamefont {V.}~\bibnamefont {Zakharova}}, \bibinfo {author} {\bibfnamefont {X.}~\bibnamefont {Davoine}}, \bibinfo {author} {\bibfnamefont {L.}~\bibnamefont {Gremillet}}, \bibinfo {author} {\bibfnamefont {M.}~\bibnamefont {Tamburini}},\ and\ \bibinfo {author} {\bibfnamefont {S.}~\bibnamefont {Corde}},\ }\href {https://doi.org/10.1038/s42005-023-01263-4} {\bibfield  {journal} {\bibinfo  {journal} {Commun. Phys.}\ }\textbf {\bibinfo {volume} {6}},\ \bibinfo {pages} {154} (\bibinfo {year} {2023})}\BibitemShut {NoStop}%
\bibitem [{\citenamefont {Yakimenko}\ \emph {et~al.}(2019{\natexlab{a}})\citenamefont {Yakimenko}, \citenamefont {Meuren}, \citenamefont {Del~Gaudio}, \citenamefont {Baumann}, \citenamefont {Fedotov}, \citenamefont {Fiuza}, \citenamefont {Grismayer}, \citenamefont {Hogan}, \citenamefont {Pukhov}, \citenamefont {Silva},\ and\ \citenamefont {White}}]{YakimenkoPRL}%
  \BibitemOpen
  \bibfield  {author} {\bibinfo {author} {\bibfnamefont {V.}~\bibnamefont {Yakimenko}}, \bibinfo {author} {\bibfnamefont {S.}~\bibnamefont {Meuren}}, \bibinfo {author} {\bibfnamefont {F.}~\bibnamefont {Del~Gaudio}}, \bibinfo {author} {\bibfnamefont {C.}~\bibnamefont {Baumann}}, \bibinfo {author} {\bibfnamefont {A.}~\bibnamefont {Fedotov}}, \bibinfo {author} {\bibfnamefont {F.}~\bibnamefont {Fiuza}}, \bibinfo {author} {\bibfnamefont {T.}~\bibnamefont {Grismayer}}, \bibinfo {author} {\bibfnamefont {M.~J.}\ \bibnamefont {Hogan}}, \bibinfo {author} {\bibfnamefont {A.}~\bibnamefont {Pukhov}}, \bibinfo {author} {\bibfnamefont {L.~O.}\ \bibnamefont {Silva}},\ and\ \bibinfo {author} {\bibfnamefont {G.}~\bibnamefont {White}},\ }\href {https://doi.org/10.1103/PhysRevLett.122.190404} {\bibfield  {journal} {\bibinfo  {journal} {Phys. Rev. Lett.}\ }\textbf {\bibinfo {volume} {122}},\ \bibinfo {pages} {190404} (\bibinfo {year} {2019}{\natexlab{a}})}\BibitemShut {NoStop}%
\bibitem [{\citenamefont {Tamburini}\ and\ \citenamefont {Meuren}(2021)}]{TamburiniPRD}%
  \BibitemOpen
  \bibfield  {author} {\bibinfo {author} {\bibfnamefont {M.}~\bibnamefont {Tamburini}}\ and\ \bibinfo {author} {\bibfnamefont {S.}~\bibnamefont {Meuren}},\ }\href {https://doi.org/10.1103/PhysRevD.104.L091903} {\bibfield  {journal} {\bibinfo  {journal} {Phys. Rev. D}\ }\textbf {\bibinfo {volume} {104}},\ \bibinfo {pages} {L091903} (\bibinfo {year} {2021})}\BibitemShut {NoStop}%
\bibitem [{\citenamefont {Corde}\ \emph {et~al.}(2020)\citenamefont {Corde}, \citenamefont {Gilljohann}, \citenamefont {Davoine}, \citenamefont {Gremillet}, \citenamefont {Sampath},\ and\ \citenamefont {Tamburini}}]{corde2020HAL}%
  \BibitemOpen
  \bibfield  {author} {\bibinfo {author} {\bibfnamefont {S.}~\bibnamefont {Corde}}, \bibinfo {author} {\bibfnamefont {M.}~\bibnamefont {Gilljohann}}, \bibinfo {author} {\bibfnamefont {X.}~\bibnamefont {Davoine}}, \bibinfo {author} {\bibfnamefont {L.}~\bibnamefont {Gremillet}}, \bibinfo {author} {\bibfnamefont {A.}~\bibnamefont {Sampath}},\ and\ \bibinfo {author} {\bibfnamefont {M.}~\bibnamefont {Tamburini}},\ }\href {https://polytechnique.hal.science/hal-02937777v2} {\emph {\bibinfo {title} {Beam focusing by near-field transition radiation}}},\ \bibinfo {type} {Research Report}\ \bibinfo {number} {hal-02937777v2}\ (\bibinfo {year} {2020})\BibitemShut {NoStop}%
\bibitem [{\citenamefont {Yakimenko}\ \emph {et~al.}(2019{\natexlab{b}})\citenamefont {Yakimenko}, \citenamefont {Alsberg}, \citenamefont {Bong}, \citenamefont {Bouchard}, \citenamefont {Clarke}, \citenamefont {Emma}, \citenamefont {Green}, \citenamefont {Hast}, \citenamefont {Hogan}, \citenamefont {Seabury}, \citenamefont {Lipkowitz}, \citenamefont {O'Shea}, \citenamefont {Storey}, \citenamefont {White},\ and\ \citenamefont {Yocky}}]{Yakimenko2019FACETII}%
  \BibitemOpen
  \bibfield  {author} {\bibinfo {author} {\bibfnamefont {V.}~\bibnamefont {Yakimenko}}, \bibinfo {author} {\bibfnamefont {L.}~\bibnamefont {Alsberg}}, \bibinfo {author} {\bibfnamefont {E.}~\bibnamefont {Bong}}, \bibinfo {author} {\bibfnamefont {G.}~\bibnamefont {Bouchard}}, \bibinfo {author} {\bibfnamefont {C.}~\bibnamefont {Clarke}}, \bibinfo {author} {\bibfnamefont {C.}~\bibnamefont {Emma}}, \bibinfo {author} {\bibfnamefont {S.}~\bibnamefont {Green}}, \bibinfo {author} {\bibfnamefont {C.}~\bibnamefont {Hast}}, \bibinfo {author} {\bibfnamefont {M.~J.}\ \bibnamefont {Hogan}}, \bibinfo {author} {\bibfnamefont {J.}~\bibnamefont {Seabury}}, \bibinfo {author} {\bibfnamefont {N.}~\bibnamefont {Lipkowitz}}, \bibinfo {author} {\bibfnamefont {B.}~\bibnamefont {O'Shea}}, \bibinfo {author} {\bibfnamefont {D.}~\bibnamefont {Storey}}, \bibinfo {author} {\bibfnamefont {G.}~\bibnamefont {White}},\ and\ \bibinfo {author} {\bibfnamefont {G.}~\bibnamefont {Yocky}},\ }\href {https://doi.org/10.1103/PhysRevAccelBeams.22.101301}
  {\bibfield  {journal} {\bibinfo  {journal} {Phys. Rev. Accel. Beams}\ }\textbf {\bibinfo {volume} {22}},\ \bibinfo {pages} {101301} (\bibinfo {year} {2019}{\natexlab{b}})}\BibitemShut {NoStop}%
\bibitem [{\citenamefont {Storey}\ \emph {et~al.}(2024)\citenamefont {Storey}, \citenamefont {Zhang}, \citenamefont {San Miguel~Claveria}, \citenamefont {Cao}, \citenamefont {Adli}, \citenamefont {Alsberg}, \citenamefont {Ariniello}, \citenamefont {Clarke}, \citenamefont {Corde}, \citenamefont {Dalichaouch}, \citenamefont {Doss}, \citenamefont {Ekerfelt}, \citenamefont {Emma}, \citenamefont {Gerstmayr}, \citenamefont {Gessner}, \citenamefont {Gilljohann}, \citenamefont {Hast}, \citenamefont {Knetsch}, \citenamefont {Lee}, \citenamefont {Litos}, \citenamefont {Loney}, \citenamefont {Marsh}, \citenamefont {Matheron}, \citenamefont {Mori}, \citenamefont {Nie}, \citenamefont {O'Shea}, \citenamefont {Parker}, \citenamefont {White}, \citenamefont {Yocky}, \citenamefont {Zakharova}, \citenamefont {Hogan},\ and\ \citenamefont {Joshi}}]{Storey2024}%
  \BibitemOpen
  \bibfield  {author} {\bibinfo {author} {\bibfnamefont {D.}~\bibnamefont {Storey}}, \bibinfo {author} {\bibfnamefont {C.}~\bibnamefont {Zhang}}, \bibinfo {author} {\bibfnamefont {P.}~\bibnamefont {San Miguel~Claveria}}, \bibinfo {author} {\bibfnamefont {G.~J.}\ \bibnamefont {Cao}}, \bibinfo {author} {\bibfnamefont {E.}~\bibnamefont {Adli}}, \bibinfo {author} {\bibfnamefont {L.}~\bibnamefont {Alsberg}}, \bibinfo {author} {\bibfnamefont {R.}~\bibnamefont {Ariniello}}, \bibinfo {author} {\bibfnamefont {C.}~\bibnamefont {Clarke}}, \bibinfo {author} {\bibfnamefont {S.}~\bibnamefont {Corde}}, \bibinfo {author} {\bibfnamefont {T.~N.}\ \bibnamefont {Dalichaouch}}, \bibinfo {author} {\bibfnamefont {C.~E.}\ \bibnamefont {Doss}}, \bibinfo {author} {\bibfnamefont {H.}~\bibnamefont {Ekerfelt}}, \bibinfo {author} {\bibfnamefont {C.}~\bibnamefont {Emma}}, \bibinfo {author} {\bibfnamefont {E.}~\bibnamefont {Gerstmayr}}, \bibinfo {author} {\bibfnamefont {S.}~\bibnamefont {Gessner}}, \bibinfo {author} {\bibfnamefont
  {M.}~\bibnamefont {Gilljohann}}, \bibinfo {author} {\bibfnamefont {C.}~\bibnamefont {Hast}}, \bibinfo {author} {\bibfnamefont {A.}~\bibnamefont {Knetsch}}, \bibinfo {author} {\bibfnamefont {V.}~\bibnamefont {Lee}}, \bibinfo {author} {\bibfnamefont {M.}~\bibnamefont {Litos}}, \bibinfo {author} {\bibfnamefont {R.}~\bibnamefont {Loney}}, \bibinfo {author} {\bibfnamefont {K.~A.}\ \bibnamefont {Marsh}}, \bibinfo {author} {\bibfnamefont {A.}~\bibnamefont {Matheron}}, \bibinfo {author} {\bibfnamefont {W.~B.}\ \bibnamefont {Mori}}, \bibinfo {author} {\bibfnamefont {Z.}~\bibnamefont {Nie}}, \bibinfo {author} {\bibfnamefont {B.}~\bibnamefont {O'Shea}}, \bibinfo {author} {\bibfnamefont {M.}~\bibnamefont {Parker}}, \bibinfo {author} {\bibfnamefont {G.}~\bibnamefont {White}}, \bibinfo {author} {\bibfnamefont {G.}~\bibnamefont {Yocky}}, \bibinfo {author} {\bibfnamefont {V.}~\bibnamefont {Zakharova}}, \bibinfo {author} {\bibfnamefont {M.~J.}\ \bibnamefont {Hogan}},\ and\ \bibinfo {author} {\bibfnamefont {C.}~\bibnamefont
  {Joshi}},\ }\href {https://doi.org/10.1103/PhysRevAccelBeams.27.051302} {\bibfield  {journal} {\bibinfo  {journal} {Phys. Rev. Accel. Beams}\ }\textbf {\bibinfo {volume} {27}},\ \bibinfo {pages} {051302} (\bibinfo {year} {2024})}\BibitemShut {NoStop}%
\bibitem [{\citenamefont {Ginzburg}\ and\ \citenamefont {Tsytovich}(1990)}]{GinzburgTsytovich1990}%
  \BibitemOpen
  \bibfield  {author} {\bibinfo {author} {\bibfnamefont {V.~L.}\ \bibnamefont {Ginzburg}}\ and\ \bibinfo {author} {\bibfnamefont {V.~N.}\ \bibnamefont {Tsytovich}},\ }\href@noop {} {\emph {\bibinfo {title} {Transition Radiation and Transition Scattering}}}\ (\bibinfo  {publisher} {Adam Hilger},\ \bibinfo {address} {Bristol, UK},\ \bibinfo {year} {1990})\ \bibinfo {note} {translated from Russian}\BibitemShut {NoStop}%
\bibitem [{\citenamefont {Jackson}(1998)}]{jackson1998}%
  \BibitemOpen
  \bibfield  {author} {\bibinfo {author} {\bibfnamefont {J.~D.}\ \bibnamefont {Jackson}},\ }\href@noop {} {\emph {\bibinfo {title} {Classical Electrodynamics}}},\ \bibinfo {edition} {3rd}\ ed.\ (\bibinfo  {publisher} {John Wiley \& Sons},\ \bibinfo {year} {1998})\BibitemShut {NoStop}%
\bibitem [{\citenamefont {Verzilov}(2000)}]{Verzilov2000}%
  \BibitemOpen
  \bibfield  {author} {\bibinfo {author} {\bibfnamefont {V.~A.}\ \bibnamefont {Verzilov}},\ }\href {https://doi.org/10.1016/S0375-9601(00)00486-2} {\bibfield  {journal} {\bibinfo  {journal} {Phys. Lett. A}\ }\textbf {\bibinfo {volume} {273}},\ \bibinfo {pages} {135} (\bibinfo {year} {2000})}\BibitemShut {NoStop}%
\bibitem [{\citenamefont {Carron}(2000)}]{Carron2000}%
  \BibitemOpen
  \bibfield  {author} {\bibinfo {author} {\bibfnamefont {N.~J.}\ \bibnamefont {Carron}},\ }\href {https://doi.org/10.2528/PIER99080102} {\bibfield  {journal} {\bibinfo  {journal} {Prog. Electromagn. Res.}\ }\textbf {\bibinfo {volume} {28}},\ \bibinfo {pages} {147} (\bibinfo {year} {2000})}\BibitemShut {NoStop}%
\bibitem [{\citenamefont {Courant}\ and\ \citenamefont {Snyder}(1958)}]{CourantSnyder}%
  \BibitemOpen
  \bibfield  {author} {\bibinfo {author} {\bibfnamefont {E.~D.}\ \bibnamefont {Courant}}\ and\ \bibinfo {author} {\bibfnamefont {H.~S.}\ \bibnamefont {Snyder}},\ }\href {https://doi.org/10.1016/0003-4916(58)90012-5} {\bibfield  {journal} {\bibinfo  {journal} {Ann. Phys.}\ }\textbf {\bibinfo {volume} {3}},\ \bibinfo {pages} {1} (\bibinfo {year} {1958})}\BibitemShut {NoStop}%
\bibitem [{\citenamefont {Adler}(1982)}]{Adler1982}%
  \BibitemOpen
  \bibfield  {author} {\bibinfo {author} {\bibfnamefont {R.~J.}\ \bibnamefont {Adler}},\ }\href {https://cds.cern.ch/record/1107997} {\bibfield  {journal} {\bibinfo  {journal} {Part. Accel.}\ }\textbf {\bibinfo {volume} {12}},\ \bibinfo {pages} {39} (\bibinfo {year} {1982})}\BibitemShut {NoStop}%
\bibitem [{\citenamefont {Humphries}(1983)}]{Humphries1983}%
  \BibitemOpen
  \bibfield  {author} {\bibinfo {author} {\bibfnamefont {S.}~\bibnamefont {Humphries}},\ }\href {https://cds.cern.ch/record/1108004} {\bibfield  {journal} {\bibinfo  {journal} {Part. Accel.}\ }\textbf {\bibinfo {volume} {13}},\ \bibinfo {pages} {249} (\bibinfo {year} {1983})}\BibitemShut {NoStop}%
\bibitem [{\citenamefont {Humphries}\ and\ \citenamefont {Ekdahl}(1988)}]{Humphries1988}%
  \BibitemOpen
  \bibfield  {author} {\bibinfo {author} {\bibfnamefont {S.}~\bibnamefont {Humphries}}\ and\ \bibinfo {author} {\bibfnamefont {C.~B.}\ \bibnamefont {Ekdahl}},\ }\href {https://doi.org/10.1063/1.340094} {\bibfield  {journal} {\bibinfo  {journal} {J. Appl. Phys.}\ }\textbf {\bibinfo {volume} {63}},\ \bibinfo {pages} {583} (\bibinfo {year} {1988})}\BibitemShut {NoStop}%
\bibitem [{\citenamefont {Humphries}\ \emph {et~al.}(1989)\citenamefont {Humphries}, \citenamefont {Ekdahl},\ and\ \citenamefont {Woodall}}]{Humphries1989}%
  \BibitemOpen
  \bibfield  {author} {\bibinfo {author} {\bibfnamefont {S.}~\bibnamefont {Humphries}}, \bibinfo {author} {\bibfnamefont {C.}~\bibnamefont {Ekdahl}},\ and\ \bibinfo {author} {\bibfnamefont {D.~M.}\ \bibnamefont {Woodall}},\ }\href {https://doi.org/10.1063/1.101142} {\bibfield  {journal} {\bibinfo  {journal} {Appl. Phys. Lett.}\ }\textbf {\bibinfo {volume} {54}},\ \bibinfo {pages} {2195} (\bibinfo {year} {1989})}\BibitemShut {NoStop}%
\bibitem [{\citenamefont {Fernsler}\ \emph {et~al.}(1990)\citenamefont {Fernsler}, \citenamefont {Hubbard},\ and\ \citenamefont {Slinker}}]{Fernsler1990}%
  \BibitemOpen
  \bibfield  {author} {\bibinfo {author} {\bibfnamefont {R.~F.}\ \bibnamefont {Fernsler}}, \bibinfo {author} {\bibfnamefont {R.~F.}\ \bibnamefont {Hubbard}},\ and\ \bibinfo {author} {\bibfnamefont {S.~P.}\ \bibnamefont {Slinker}},\ }\href {https://doi.org/10.1063/1.346932} {\bibfield  {journal} {\bibinfo  {journal} {J. Appl. Phys.}\ }\textbf {\bibinfo {volume} {68}},\ \bibinfo {pages} {5985} (\bibinfo {year} {1990})}\BibitemShut {NoStop}%
\bibitem [{\citenamefont {Bethe}\ and\ \citenamefont {Heitler}(1934)}]{Bethe1934}%
  \BibitemOpen
  \bibfield  {author} {\bibinfo {author} {\bibfnamefont {H.}~\bibnamefont {Bethe}}\ and\ \bibinfo {author} {\bibfnamefont {W.}~\bibnamefont {Heitler}},\ }\href {https://doi.org/10.1098/rspa.1934.0140} {\bibfield  {journal} {\bibinfo  {journal} {Proc. R. Soc. London, Ser. A}\ }\textbf {\bibinfo {volume} {146}},\ \bibinfo {pages} {83} (\bibinfo {year} {1934})}\BibitemShut {NoStop}%
\bibitem [{\citenamefont {Doss}\ \emph {et~al.}(2019)\citenamefont {Doss}, \citenamefont {Adli}, \citenamefont {Ariniello}, \citenamefont {Cary}, \citenamefont {Corde}, \citenamefont {Hidding}, \citenamefont {Hogan}, \citenamefont {Hunt-Stone}, \citenamefont {Joshi}, \citenamefont {Marsh}, \citenamefont {Rosenzweig}, \citenamefont {Vafaei-Najafabadi}, \citenamefont {Yakimenko},\ and\ \citenamefont {Litos}}]{Doss2019}%
  \BibitemOpen
  \bibfield  {author} {\bibinfo {author} {\bibfnamefont {C.~E.}\ \bibnamefont {Doss}}, \bibinfo {author} {\bibfnamefont {E.}~\bibnamefont {Adli}}, \bibinfo {author} {\bibfnamefont {R.}~\bibnamefont {Ariniello}}, \bibinfo {author} {\bibfnamefont {J.}~\bibnamefont {Cary}}, \bibinfo {author} {\bibfnamefont {S.}~\bibnamefont {Corde}}, \bibinfo {author} {\bibfnamefont {B.}~\bibnamefont {Hidding}}, \bibinfo {author} {\bibfnamefont {M.~J.}\ \bibnamefont {Hogan}}, \bibinfo {author} {\bibfnamefont {K.}~\bibnamefont {Hunt-Stone}}, \bibinfo {author} {\bibfnamefont {C.}~\bibnamefont {Joshi}}, \bibinfo {author} {\bibfnamefont {K.~A.}\ \bibnamefont {Marsh}}, \bibinfo {author} {\bibfnamefont {J.~B.}\ \bibnamefont {Rosenzweig}}, \bibinfo {author} {\bibfnamefont {N.}~\bibnamefont {Vafaei-Najafabadi}}, \bibinfo {author} {\bibfnamefont {V.}~\bibnamefont {Yakimenko}},\ and\ \bibinfo {author} {\bibfnamefont {M.}~\bibnamefont {Litos}},\ }\href {https://doi.org/10.1103/PhysRevAccelBeams.22.111001} {\bibfield  {journal} {\bibinfo
  {journal} {Phys. Rev. Accel. Beams}\ }\textbf {\bibinfo {volume} {22}},\ \bibinfo {pages} {111001} (\bibinfo {year} {2019})}\BibitemShut {NoStop}%
\bibitem [{\citenamefont {Emma}\ \emph {et~al.}(2021)\citenamefont {Emma}, \citenamefont {Xu}, \citenamefont {Fisher}, \citenamefont {Robles}, \citenamefont {MacArthur}, \citenamefont {Cryan}, \citenamefont {Hogan}, \citenamefont {Musumeci}, \citenamefont {White},\ and\ \citenamefont {Marinelli}}]{Emma2021}%
  \BibitemOpen
  \bibfield  {author} {\bibinfo {author} {\bibfnamefont {C.}~\bibnamefont {Emma}}, \bibinfo {author} {\bibfnamefont {X.}~\bibnamefont {Xu}}, \bibinfo {author} {\bibfnamefont {A.}~\bibnamefont {Fisher}}, \bibinfo {author} {\bibfnamefont {R.}~\bibnamefont {Robles}}, \bibinfo {author} {\bibfnamefont {J.~P.}\ \bibnamefont {MacArthur}}, \bibinfo {author} {\bibfnamefont {J.}~\bibnamefont {Cryan}}, \bibinfo {author} {\bibfnamefont {M.~J.}\ \bibnamefont {Hogan}}, \bibinfo {author} {\bibfnamefont {P.}~\bibnamefont {Musumeci}}, \bibinfo {author} {\bibfnamefont {G.}~\bibnamefont {White}},\ and\ \bibinfo {author} {\bibfnamefont {A.}~\bibnamefont {Marinelli}},\ }\href {https://doi.org/10.1063/5.0050693} {\bibfield  {journal} {\bibinfo  {journal} {APL Photonics}\ }\textbf {\bibinfo {volume} {6}},\ \bibinfo {pages} {076107} (\bibinfo {year} {2021})}\BibitemShut {NoStop}%
\bibitem [{\citenamefont {Lehe}\ \emph {et~al.}(2016)\citenamefont {Lehe}, \citenamefont {Kirchen}, \citenamefont {Andriyash}, \citenamefont {Godfrey},\ and\ \citenamefont {Vay}}]{Lehe2016}%
  \BibitemOpen
  \bibfield  {author} {\bibinfo {author} {\bibfnamefont {R.}~\bibnamefont {Lehe}}, \bibinfo {author} {\bibfnamefont {M.}~\bibnamefont {Kirchen}}, \bibinfo {author} {\bibfnamefont {I.~A.}\ \bibnamefont {Andriyash}}, \bibinfo {author} {\bibfnamefont {B.~B.}\ \bibnamefont {Godfrey}},\ and\ \bibinfo {author} {\bibfnamefont {J.-L.}\ \bibnamefont {Vay}},\ }\href {https://doi.org/10.1016/j.cpc.2016.02.007} {\bibfield  {journal} {\bibinfo  {journal} {Comput. Phys. Commun.}\ }\textbf {\bibinfo {volume} {203}},\ \bibinfo {pages} {66} (\bibinfo {year} {2016})}\BibitemShut {NoStop}%
\bibitem [{\citenamefont {Matheron}(2025)}]{MatheronThese}%
  \BibitemOpen
  \bibfield  {author} {\bibinfo {author} {\bibfnamefont {A.}~\bibnamefont {Matheron}},\ }\emph {\bibinfo {title} {{Extreme plasma interactions for Strong-Field QED}}},\ \href {https://theses.hal.science/tel-05107259} {Ph.D. thesis},\ \bibinfo  {school} {{Institut Polytechnique de Paris}} (\bibinfo {year} {2025})\BibitemShut {NoStop}%
\bibitem [{\citenamefont {Loos}\ \emph {et~al.}(2007)\citenamefont {Loos}, \citenamefont {Borden}, \citenamefont {Emma}, \citenamefont {Frisch},\ and\ \citenamefont {Wu}}]{Loos2007}%
  \BibitemOpen
  \bibfield  {author} {\bibinfo {author} {\bibfnamefont {H.}~\bibnamefont {Loos}}, \bibinfo {author} {\bibfnamefont {T.}~\bibnamefont {Borden}}, \bibinfo {author} {\bibfnamefont {P.}~\bibnamefont {Emma}}, \bibinfo {author} {\bibfnamefont {J.}~\bibnamefont {Frisch}},\ and\ \bibinfo {author} {\bibfnamefont {J.}~\bibnamefont {Wu}},\ }in\ \href {https://proceedings.jacow.org/p07/PAPERS/FRPMS071.PDF} {\emph {\bibinfo {booktitle} {Proceedings of Particle Accelerator Conference (PAC 2007)}}}\ (\bibinfo {year} {2007})\ p.\ \bibinfo {pages} {4189},\ \bibinfo {note} {paper FRPMS071}\BibitemShut {NoStop}%
\bibitem [{\citenamefont {Lehe}\ \emph {et~al.}(2024)\citenamefont {Lehe}, \citenamefont {Kirchen}, \citenamefont {Jalas}, \citenamefont {Jeppe}, \citenamefont {Andriyash}, \citenamefont {Peters}, \citenamefont {Huebl}, \citenamefont {Yoffe}, \citenamefont {Seemann}, \citenamefont {Dornmair}, \citenamefont {Poder}, \citenamefont {de~la Ossa}, \citenamefont {Zoni}, \citenamefont {Grote}, \citenamefont {Sta{\'n}czak-Marikin}, \citenamefont {Shalloo}, \citenamefont {Golovanov}, \citenamefont {Isaiah}, \citenamefont {Th{\'e}venet}, \citenamefont {Pausch}, \citenamefont {Kuschel}, \citenamefont {Seipt}, \citenamefont {Eurazov},\ and\ \citenamefont {Hui}}]{FBPICGithub}%
  \BibitemOpen
  \bibfield  {author} {\bibinfo {author} {\bibfnamefont {R.}~\bibnamefont {Lehe}}, \bibinfo {author} {\bibfnamefont {M.}~\bibnamefont {Kirchen}}, \bibinfo {author} {\bibfnamefont {S.}~\bibnamefont {Jalas}}, \bibinfo {author} {\bibfnamefont {L.}~\bibnamefont {Jeppe}}, \bibinfo {author} {\bibfnamefont {I.}~\bibnamefont {Andriyash}}, \bibinfo {author} {\bibfnamefont {K.}~\bibnamefont {Peters}}, \bibinfo {author} {\bibfnamefont {A.}~\bibnamefont {Huebl}}, \bibinfo {author} {\bibfnamefont {S.}~\bibnamefont {Yoffe}}, \bibinfo {author} {\bibfnamefont {O.}~\bibnamefont {Seemann}}, \bibinfo {author} {\bibfnamefont {M.}~\bibnamefont {Dornmair}}, \bibinfo {author} {\bibfnamefont {K.}~\bibnamefont {Poder}}, \bibinfo {author} {\bibfnamefont {A.}~\bibnamefont {de~la Ossa}}, \bibinfo {author} {\bibfnamefont {E.}~\bibnamefont {Zoni}}, \bibinfo {author} {\bibfnamefont {D.}~\bibnamefont {Grote}}, \bibinfo {author} {\bibfnamefont {D.}~\bibnamefont {Sta{\'n}czak-Marikin}}, \bibinfo {author} {\bibfnamefont {R.}~\bibnamefont
  {Shalloo}}, \bibinfo {author} {\bibfnamefont {A.}~\bibnamefont {Golovanov}}, \bibinfo {author} {\bibnamefont {Isaiah}}, \bibinfo {author} {\bibfnamefont {M.}~\bibnamefont {Th{\'e}venet}}, \bibinfo {author} {\bibfnamefont {R.}~\bibnamefont {Pausch}}, \bibinfo {author} {\bibfnamefont {S.}~\bibnamefont {Kuschel}}, \bibinfo {author} {\bibfnamefont {D.}~\bibnamefont {Seipt}}, \bibinfo {author} {\bibnamefont {Eurazov}},\ and\ \bibinfo {author} {\bibfnamefont {D.-X.}\ \bibnamefont {Hui}},\ }\href {https://doi.org/10.5281/zenodo.12824134} {\bibinfo {title} {{fbpic/fbpic}: {0.26.1}}},\ \bibinfo {howpublished} {Zenodo} (\bibinfo {year} {2024})\BibitemShut {NoStop}%
\end{thebibliography}
\end{document}